\documentclass{jfm} 

\usepackage{graphicx}
\usepackage{amsmath}
\usepackage{amssymb}
\usepackage{color}
\usepackage{natbib}
\usepackage{hyperref}
\usepackage{caption}
\usepackage{epstopdf, epsfig}

\graphicspath{{figures/}}

\shorttitle{Distributed flexibility in inertial swimmers}
\shortauthor{D. Floryan and C. W. Rowley}

\title{Distributed flexibility in inertial swimmers}

\author{Daniel Floryan\aff{1}
  \corresp{\email{dfloryan@alumni.princeton.edu}}
 \and Clarence W. Rowley\aff{1}}

\affiliation{\aff{1}Department of Mechanical and Aerospace Engineering, Princeton University,
Princeton, NJ 08544, USA}

\begin{document}

\maketitle

\begin{abstract}

We study a linear inviscid model of a passively flexible swimmer with distributed flexibility, calculating its propulsive performance and optimal distributions of flexibility. The frequencies of actuation and mean stiffness ratios we consider span a large range, while the mass ratio is fixed to a low value representative of swimmers. We present results showing how the trailing edge deflection, thrust coefficient, power coefficient, and efficiency vary with frequency, mean stiffness, and stiffness distribution. Swimmers with distributed flexibility have the same qualitative features as those with uniform flexibility. Significant gains in thrust can be made, however, by tuning the stiffness such that a resonant response is triggered, or by concentrating stiffness towards the leading edge if resonance cannot be triggered. To minimize power, the opposite is true. Meaningful gains in efficiency can be made at low frequencies by concentrating stiffness away from the leading edge, since doing so induces efficient travelling wave kinematics. We also speculate on the effects of a finite Reynolds number in the form of streamwise drag. The drag adds an offset to the net thrust produced by the swimmer, causing efficiency-maximizing distributions of flexibility to tend towards thrust-maximizing ones, representative of what is found in nature. 

\end{abstract}


\section{Introduction}
\label{sec:intro}

Animal swimming and flight are complex. To make headway in understanding how
animals swim and fly, we often abstract the coordinated motion of entire bodies
to plates flapping in a fluid; a vast literature studies this simplified
canonical problem (see the reviews in \citet{triantafyllou2000hydrodynamics}, \citet{wang2005dissecting} and \citet{wu2011fish}, for example). A salient feature of the fins
and wings of swimming and flying animals is flexibility. Flexibility allows fish
to use fine musculature to actively control their kinematics to some degree
\citep{fish2006passive}, and allows birds to morph their wings
\citep{bergmann1839bewegungen}, but also passively influences kinematics through
elastic restoring forces.

Simple flapping plate models of swimming and flight incorporate flexibility by
modelling the plate as a uniformly elastic material, allowing it to deform
according to the fluid and elastic forces it experiences. In the context of
forward propulsion, we are most interested in the thrust that a flapping plate
can produce, as well as how efficiently it produces the thrust. Passive
flexibility changes the thrust that a flapping plate produces, as well as the
efficiency of thrust production. It has generally been found that, compared to
rigid plates, uniformly flexible plates produce greater thrust when actuated
near a fluid-structure natural frequency, and less thrust otherwise, but the
efficiency of uniformly flexible plates is greater than that of rigid plates
over a broad range of frequencies and stiffnesses \citep{alben2008optimal,
  de2011thrust, dewey2013scaling, katz1978hydrodynamic, katz1979large,
  quinn2014scaling, floryan2018clarifying}. While thrust generally exhibits
local maxima when actuating near natural frequencies, efficiency has been
observed to exhibit local maxima below natural frequencies, near natural
frequencies, and above natural frequencies \citep{dewey2013scaling,
  moored2014linear, quinn2014scaling, quinn2015maximizing, paraz2016thrust}, as
well as at frequencies relatively far from a natural frequency
\citep{ramananarivo2011rather, kang2011effects, vanella2009influence,
  zhu2014flexibility, michelin2009resonance}. We recently clarified that
resonant behaviour in efficiency---at least for swimmers, where the
characteristic fluid mass is much greater than the body mass---can arise only
when viscous forces are present, or if nonlinear effects are not negligible
\citep{floryan2018clarifying}.

The above studies consider plates for which the stiffness is {\em uniform} along
the chord; however, the flexibility of fins and wings of real animals is
typically non-uniform. The material properties of fins and wings may change along
the chord (as the musculature, fat content, and skin changes, for example), as
may the thickness. (Figure~17 in \citet{fish2006passive} shows a beautiful
example of varying material properties and thickness of the fluke of a
bottlenose dolphin.) Flexibility may even be highly localized, as in the veined
wings of insects \citep{combes2003bflexural}. We thus ask how \emph{distributed}
(non-uniform, heterogeneous) flexibility affects thrust production and efficiency
in flapping plates, in contrast to uniform flexibility.

Only recently have people begun to explore how the distribution of flexibility
affects propulsion in flapping plates. Experiments tend to focus on biomimetic
flexibility distributions similar to fish fins, where the leading portion of the
plate is stiffer than the trailing portion. The literature includes results on
distributions that are fully biomimetic with pure pitching motions
\citep{riggs2010advantages}, stepwise constant distributions with pure heaving
and zero angle of attack motions \citep{lucas2015effects}, and supposedly linear
distributions with pure heaving motions \citep{kancharala2016optimal}; all of
these experiments were for cases where the characteristic fluid mass is much
greater than the characteristic body mass, as in swimmers. The experiments
generally show that plates that are stiffer towards the leading edge produce
more thrust and do so more efficiently than plates that are uniformly
flexible. It is very important to note, however, that in the cited works, the
plates with uniform and distributed flexibilities had different mean
stiffnesses, making it difficult to distinguish between the effects of changes
in mean stiffness and changes in stiffness distribution. Being able to make the
distinction is important because, as we will show later, changing the mean
stiffness can significantly change natural frequencies, which have significant
effects on thrust and efficiency, and changing the mean stiffness also changes
the off-resonance behaviour in efficiency \citep{alben2008optimal,
  floryan2018clarifying}.

Computational works have also analysed how the distribution of flexibility
affects propulsion. In most studies, the characteristic fluid mass is of the
same order as the characteristic body mass, as in fliers (many of these studies
are motivated by insect flight). Distributed flexibility has been modelled in
several ways: as a uniform elastic plate with virtual linear springs at several
control points (the virtual linear springs attach the elastic sheet to points
with \textit{a priori} known motions, mimicking veins in insect wings)
\citep{shoele2013performance}; as an elastic plate with varying material
properties \citep{moore2015torsional}; and as an elastic plate with homogeneous
material properties but varying thickness \citep{yeh2017efficient}. Both
\citet{shoele2013performance} and \citet{yeh2017efficient} found that plates
with stiff leading edges produced thrust curves that had lower, but broader,
peaks than those of uniformly flexible plates, and that plates with stiff
leading edges were broadly more efficient than uniformly flexible
plates. \citet{moore2015torsional} optimized the stiffness (mean and
distribution) at fixed frequencies for thrust, and found that a plate that is
rigid except at the leading edge (where it has a torsional spring) produced
greater thrust than any other flexible plate (although the thrust is not much
greater than that produced by a plate with linearly distributed
flexibility). The only computational work directly applicable to swimmers, for which
the characteristic fluid mass is much greater than the characteristic body mass,
is \citet{kancharala2016optimal}, where the authors found that a stiffer leading
edge enhances thrust and efficiency for their kinematics.

Although several studies have shown that distributed flexibility can enhance the
propulsion of flexible flapping plates in some way, the mechanisms are
unclear. In particular, none of the studies mentioned above have controlled for mean
stiffness, which is known to significantly affect propulsion, so it is
impossible to know how the distribution of flexibility alone affects
propulsion. The eigenvalues of a suitable linear system can provide a basis to
understand how the distribution of flexibility affects propulsion, but this
approach has not yet been pursued. Furthermore, the literature has only given
conditions that, if met, give rise to improvements in thrust or efficiency, but
this is far from a complete characterization of the effects of the distribution
of flexibility. For example, although we know that a plate with a stiff leading
edge operating at a certain frequency and mean stiffness produces greater thrust
than a uniformly flexible plate with a different mean stiffness, we cannot
generalize this statement to other cases, or conclude that other distributions
do not improve propulsion.

In this work, we attempt to characterize how distributed flexibility, in contrast to uniform flexibility, changes the thrust production, power consumption, and efficiency of propulsion of flapping plates. We emphasize the role of the distribution of flexibility---separate from its mean value---particularly how it alters natural frequencies and resonance. We also calculate optimal stiffness distributions, and explain them in light of the preceding analysis. To be clear, our own interests lie mainly in inertial swimmers characterized by high Reynolds numbers and a large ratio of characteristic fluid mass to body mass. This is in contrast to fliers, for example, where the mass ratio is of order unity and higher. We employ a linear model of a passively flexible swimmer, since doing so allows us to formally calculate natural frequencies of the coupled fluid-structure system, and to stay in a dynamical regime where the notion of resonance is clear. 

For finite-amplitude swimmers, nonlinear effects may be important. The results of \citet{michelin2009resonance} suggest that the main effect of nonlinearity is to effectively increase damping in the system, as resonant peaks weaken, broaden, and shift to lower frequencies with increasing amplitude of actuation. This effect is verified in \citet{goza2020connections}, where the authors explicitly delineate linear and nonlinear effects with increasing amplitude. In addition, the authors find that for large enough amplitudes, leading edge separation may occur, which can drastically alter the performance characteristics from what one would expect through linear predictions, echoing the results of \citet{quinn2015maximizing}. Therefore, our linear results should extend qualitatively to finite-amplitude swimmers as long as leading edge separation is avoided. For more details, we refer the reader to \citet{goza2020connections}.

\section{Problem description}
\label{sec:prob}
\begin{figure}
  \begin{center}
  \includegraphics[width=\linewidth]{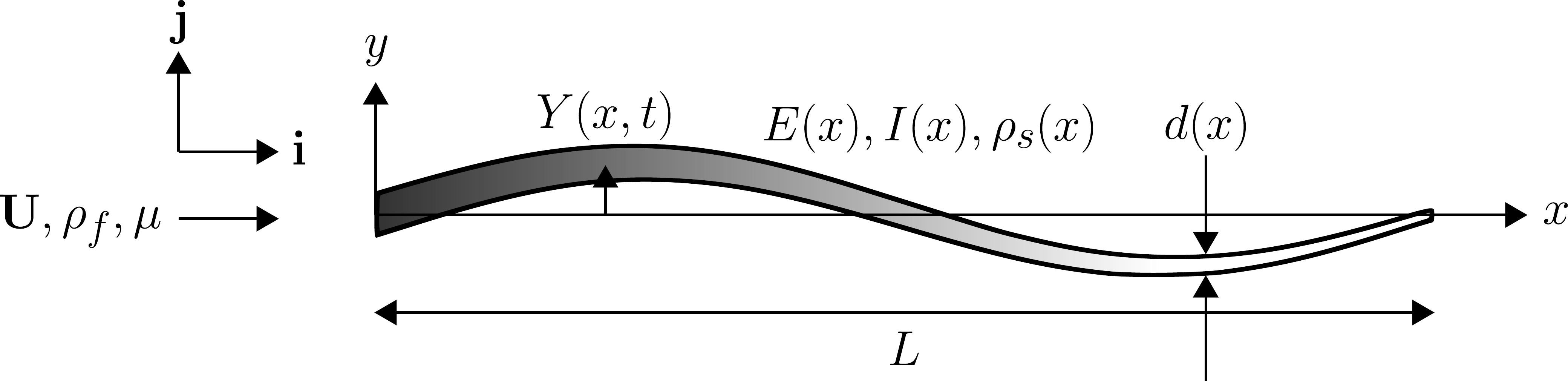}
  \end{center}
  \caption{Schematic of the problem. The varying colour represents the varying material properties. }
  \label{fig:setup}
\end{figure}

Here, the set-up and assumptions are the same as in \citet{moore2017fast}. Consider a two-dimensional, inextensible elastic plate of length~$L$ and thickness~$d$. The plate is thin (${d \ll L}$), and is transversely deflected a small amount $Y$ from its neutral position, with its slope ${Y_x \ll 1}$. Under these assumptions, the dynamics of the plate is governed by Euler-Bernoulli beam theory. The plate has density~$\rho_s$ and flexural rigidity~$B = EI$, where $E$ is the Young's modulus, $I = wd^3/12$ is the second moment of area of the plate, and $w$ is the width of the plate. We allow the properties of the plate to \emph{vary spatially}; that is, $\rho_s$, $E$, and $d$ are functions of $x$. The plate is immersed in an incompressible, inviscid Newtonian fluid of density~$\rho_f$. There is no flow along the width of the plate, and far from the plate the flow is unidirectional and constant: $\mathbf{U} = U\mathbf{i}$. The set-up is altogether illustrated in figure~\ref{fig:setup}. 

The motion of the plate alters the velocity field of the fluid, whose forces in turn modify the motion of the plate. The transverse position of the plate satisfies the Euler-Bernoulli beam equation
\begin{equation}
  \label{eq:prob1}
  \rho_s d w Y_{tt} + (B Y_{xx})_{xx} = w \Delta p,
\end{equation}
where $\Delta p$ is the pressure difference across the plate due to the fluid flow, subscript $t$ denotes differentiation with respect to time, and subscript $x$ denotes differentiation with respect to streamwise position. The fluid motion satisfies the linearized incompressible Euler equations
\begin{equation}
  \label{eq:prob2}
  \left. \begin{array}{ll}
  \nabla \cdot \mathbf{u} = 0, \\[8pt]
  \rho_f(\mathbf{u}_t + U\mathbf{u}_x) = -\nabla p,
  \end{array}\right\}
\end{equation}
where $\mathbf{u} = u\mathbf{i} + v\mathbf{j}$. The above linearization is valid when the perturbation velocity $\mathbf{u}$ is much smaller than $U$. Since the perturbation velocity depends on the plate's vertical velocity, its slope, and the rate of change of its slope, the linear assumption holds for small-amplitude motions of the plate. 

We non-dimensionalize the above equations using $L/2$ as the length scale, $U$ as the velocity scale, and $L/(2U)$ as the time scale, yielding
\begin{equation}
  \label{eq:prob3}
  \left. \begin{array}{ll}
  2R Y_{tt} + \displaystyle\frac{2}{3}(S Y_{xx})_{xx} = \Delta p, \\[8pt]
  \nabla \cdot \mathbf{u} = 0, \\[8pt]
  \mathbf{u}_t + \mathbf{u}_x = \nabla \phi, 
  \end{array}\right\}
\end{equation}
where
\begin{equation}
  \label{eq:prob4}
  R(x) = \frac{\rho_s d(x)}{\rho_f L}, \quad S(x) = \frac{E(x) d(x)^3}{\rho_f U^2 L^3}, \quad \phi = p_\infty - p.
\end{equation}
In the above, $x$, $t$, $Y$, $\mathbf{u}$, and $p$ are now dimensionless, with
the pressure non-dimensionalized by~$\rho_f U^2$.
The coordinates are aligned
such that $x = -1$ corresponds to the leading edge and $x = 1$ corresponds to
the trailing edge; $R$ is a ratio of solid-to-fluid mass, and $S$ is a ratio of
bending-to-fluid forces, and both are functions of $x$. Note that $\Delta \phi =
-\Delta p$.

The fluid additionally satisfies the no-penetration and Kutta conditions, which
for small-amplitude motions take the form 
\begin{equation}
  \label{eq:prob5}
  \left. \begin{array}{ll}
  v|_{x \in [-1,1], y = 0} = Y_t + Y_x, \\[8pt]
  |v||_{(x,y)=(1,0)} < \infty.
  \end{array}\right\}
\end{equation}

We impose heaving and pitching motions $h$ and $\theta$, respectively, on the leading edge of the plate, while the trailing edge is free, resulting in boundary conditions
\begin{equation}
  \label{eq:prob6}
  Y(-1,t) = h(t),\quad Y_x(-1,t) = \theta(t),\quad Y_{xx}(1,t) = 0,\quad Y_{xxx}(1,t) = 0. 
\end{equation}
The fluid motion resulting from the actuation of the leading edge of the plate imparts a net horizontal force onto the plate. In other words, energy input into the system by the actuation of the leading edge is used to generate a propulsive force. The net horizontal force (thrust) on the plate is
\begin{equation}
  \label{eq:prob7}
  C_T = \int_{-1}^1 (\Delta p) Y_x \, \mathrm{d}x + C_{TS},
\end{equation}
where $C_{TS}$ is the leading edge suction force given by
\begin{equation}
  \label{eq:suction}
  C_{TS} = \frac{\upi}{32} \lim_{x \to -1} \gamma^2 (1 - x^2).
\end{equation}
Above, $\gamma$ is the jump in tangential velocity across the plate, i.e., the bound vortex sheet strength \citep{alben2009simulating}. Ultimately, we use the formula given in \cite{moore2017fast}. The power input is
\begin{equation}
  \label{eq:prob8}
  C_P = -\int_{-1}^1 (\Delta p) Y_t \, \mathrm{d}x.
\end{equation}
The leading edge suction force used is the limit of the
suction force on a leading edge of small but finite radius of curvature, in the
limit that the radius tends to zero. The leading edge suction force is a
reasonable model of the actual flow when it is attached
\citep{saffman1992vortex}, so we have chosen to include
it. In terms of dimensional variables, $C_T = T/(\frac{1}{2}\rho_f U^2 Lw)$ and
$C_P = P/(\frac{1}{2}\rho_f U^3 Lw)$, where $T$ and $P$ are the dimensional net
thrust and power input, respectively. Finally, the Froude efficiency is defined
as
\begin{equation}
  \label{eq:prob12}
  \eta = \frac{\overline{T} U}{\overline{P}} = \frac{\overline{C_T}}{\overline{C_P}},
\end{equation}
where the overbar denotes a time-averaged quantity. 

In this work, we restrict ourselves to actuation at the leading edge that is sinusoidal in time, that is, 
\begin{equation}
  \label{eq:prob13}
  \left. \begin{array}{ll}
  h(t) = \Real\{h_0 \mathrm{e}^{\mathrm{i}\sigma t}\}, \\[8pt]
  \theta(t) = \Real\{\theta_0 \mathrm{e}^{\mathrm{i}\sigma t}\},
  \end{array}\right\}
\end{equation}
where $\sigma = \upi Lf/U$ is the dimensionless angular frequency, $f$ is the
dimensional frequency in Hz, $\mathrm{i} = \sqrt{-1}$, and $\Real$ denotes the 
real part of a complex number. Since the system is
linear in $Y$, the resulting deflection of the plate and fluid flow will also be
sinusoidal in time.
We leave the details of the method of solution to Appendix~\ref{sec:sol}, noting
that all calculations in this work used either 64 or 128 collocation points. Low values of mean stiffness coupled with high actuation frequencies lead to deflection patterns with short wavelengths that are not adequately resolved by the number of collocation points used; we have been careful to make note of and remove such cases from the presented results. The
method to calculate the eigenvalues of the system is detailed in
Appendix~\ref{sec:eig}, and some useful formulas for the numerical method used
are given in Appendix~\ref{sec:form}.

\section{Parameters and scope}
\label{sec:param}
The system parameters we use will critically affect the phenomena we observe. We thus take the opportunity here to explicitly state the parameters we use in this work, noting some attendant qualitative features. 

The system is parameterized by its Reynolds number, $Re$, mass (mean and distribution), stiffness (mean and distribution), and frequency and amplitude of actuation. Our flow is inviscid, but we will briefly remark on the effects of a finite Reynolds number later. The non-dimensional quantities in \eqref{eq:prob4} show that the mass and stiffness of the system depend on both the solid \emph{and} the fluid. Underwater swimmers tend to be thin and neutrally buoyant, so the mass ratio $R$ is generally quite low; this is in contrast to fliers, for example, whose mass ratios are of order unity and higher. Since our interests lie in swimmers, we take the mean mass ratio to be $\langle R \rangle = 0.01$ throughout, where $\langle \cdot \rangle$ denotes the spatial mean along the length of the plate. The stiffness of the system is characterized by the stiffness ratio $S$; we vary the mean stiffness of the system from very flexible ($\langle S \rangle \ll 1$) to very stiff ($\langle S \rangle \gg 1$). We vary the frequency of actuation so that it covers multiple natural frequencies of the system. Since our system is linear, scaling the amplitude by some factor will simply scale the flow and deflection fields by the same factor. In this sense, amplitude does not matter in our problem, so we set the heaving and pitching amplitudes so that the maximum deflection of the trailing edge of a rigid plate is equal to the length of the plate. The amplitude affects both thrust and power quadratically, and does not affect efficiency in this linear setting. We do not consider nonlinear effects caused by large amplitudes. Since we study an infinitesimally thin plate, we do not consider any geometric effects, which could reasonably be argued to be important. The parameters we use in the following sections are summarized in Table~\ref{tab:param}. 

\begin{table}
  \begin{center}
\def~{\hphantom{0}}
  \begin{tabular}{cccccc} \hline
      $Re$ & $\langle R \rangle = \displaystyle\frac{\langle \rho_s d \rangle}{\rho_f L}$ & $\langle S \rangle = \displaystyle\frac{\langle Ed^3 \rangle}{\rho_f U^2 L^3}$ & $f^* = \displaystyle\frac{fL}{U}$ & $h_0$ & $\theta_0$ \\[8pt]
       inviscid & $0.01$ & $10^{-2}$--$10^2$ & $10^{-1}$--$10^2$ & 2 (linear) & 1 (linear) \\ \hline
  \end{tabular}
  \caption{Parameter values used in this work.}
  \label{tab:param}
  \end{center}
\end{table}

\begin{figure}
  \begin{center}
  \includegraphics[width=\linewidth]{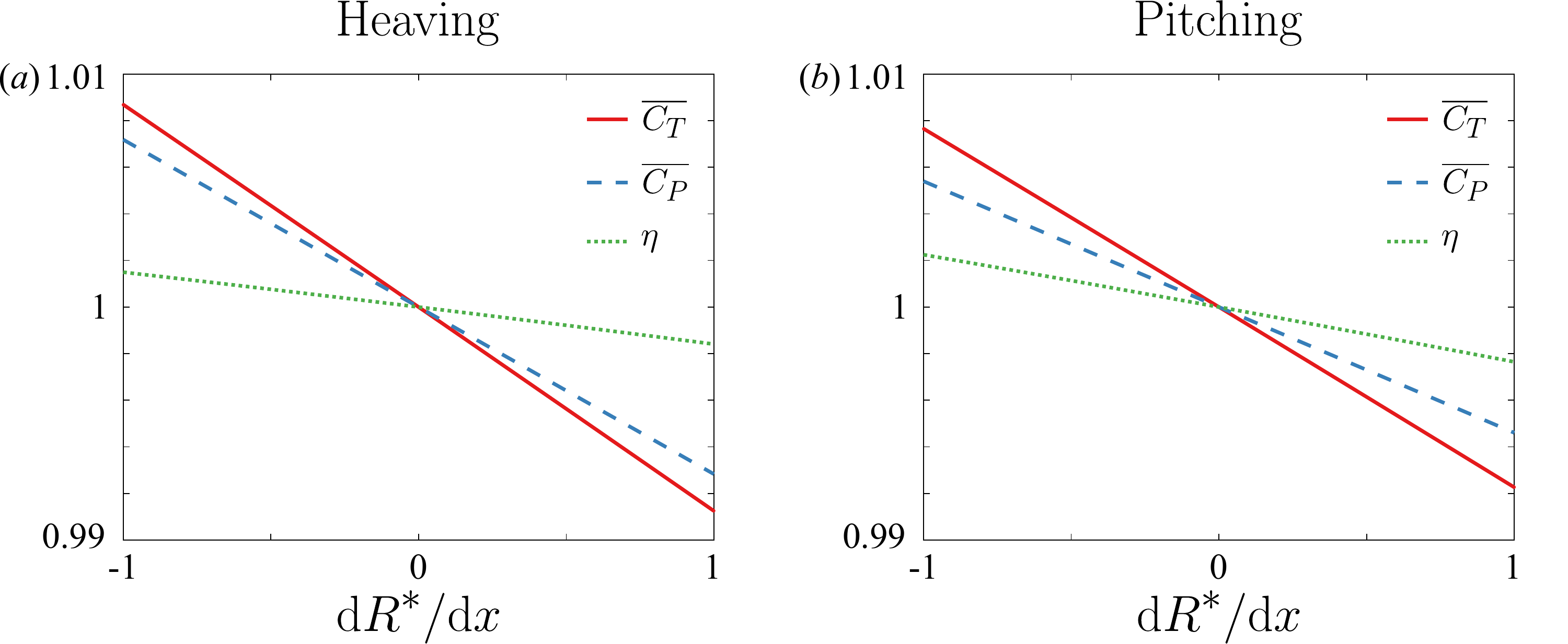}
  \end{center}
  \caption{Thrust coefficient, power coefficient, and efficiency as a function
    of the mass distribution for a (a) heaving and (b) pitching plate for
    $\langle R \rangle = 0.01$, $S \equiv 1$, and $f^* = 1$. The mass
    distribution~$R$ is linear, and values are normalized by their value when
    the mass is uniformly distributed.}
  \label{fig:mass}
\end{figure}

\begin{figure}
  \begin{center}
  \includegraphics[width=\linewidth]{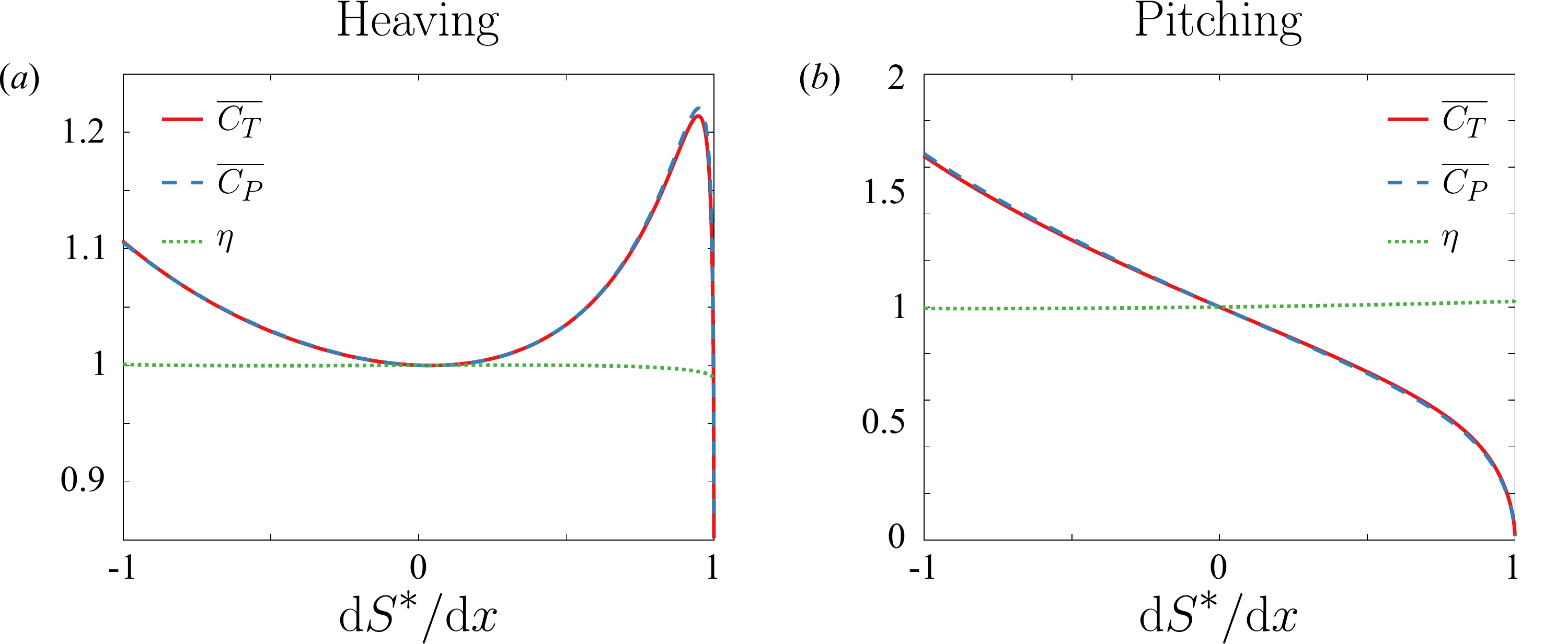}
  \end{center}
  \caption{Thrust coefficient, power coefficient, and efficiency as a function
    of the stiffness distribution for a (a) heaving and (b) pitching plate for
    $R \equiv 0.01$, $\langle S \rangle = 1$, and $f^* = 1$. The stiffness
    distribution~$S$ is linear, and values are normalized by their value when
    the mass is uniformly distributed.}
  \label{fig:stiffness}
\end{figure}

As shown in \citet{floryan2018clarifying}, the value of the mean mass ratio
$\langle R \rangle$ qualitatively changes the propulsion of a flapping plate. At
low values, however, the mass of the plate is dominated by the mass of the
fluid. With $\langle R \rangle = 0.01$, we expect the mass of the plate to have
little effect on propulsion, and consequently the distribution of mass should
also have little effect on propulsion, at least for cases where there is not a
large amount of mass concentrated in a small area. In figure~\ref{fig:mass}, we
plot the thrust coefficient, power coefficient, and efficiency as functions of
the mass distribution for $\langle R \rangle = 0.01$, $S \equiv 1$, and $f^* =
1$ for heaving and pitching plates. Here, we have taken the mass to be
distributed linearly, in which case it is described by a single parameter
$\text{d}R^*/\text{d}x$, where $R = \langle R \rangle R^*$, $R^*$ is the
distribution of mass, and hence $\langle R^* \rangle = 1$. Note that
$\text{d}R^*/\text{d}x \in [-1,1]$ (otherwise a section of the plate would have
negative mass), where $\text{d}R^*/\text{d}x = -1$ corresponds to a massive
leading edge, $\text{d}R^*/\text{d}x = 0$ corresponds to uniformly distributed
mass, and $\text{d}R^*/\text{d}x = 1$ corresponds to a massive trailing
edge. The thrust coefficient, power coefficient, and efficiency in
figure~\ref{fig:mass} have been normalized by their values when the mass is
uniformly distributed. At such low $\langle R \rangle$ the distribution of mass
matters little; this is in contrast to stiffness, whose distribution can greatly
affect thrust, power, and efficiency, as shown in figure~\ref{fig:stiffness} for
the same mean parameter values as used in figure~\ref{fig:mass}. Accordingly, we
will take $R \equiv 0.01$ in all results below. We expect our results to hold for low mass ratios ($R \lesssim 0.1$). We will allow the stiffness to vary in space, with the mean value given by $\langle S \rangle$ and the distribution given by~$S^*$. The distribution~$S^*$ can be conveniently described as a linear combination of Legendre polynomials, 
\begin{equation}
  \label{eq:param1}
  S^* = \sum_i c_i P_i,
\end{equation}
the first few of which are
\begin{equation}
  \label{eq:param2}
  P_0(x) = 1, \qquad P_1(x) = x, \qquad P_2(x) = \frac{1}{2}(3x^2 - 1).
\end{equation}
Legendre polynomials are convenient because they are orthogonal on $x \in [-1,1]$ with weighting function 1. Consequently, we can fix $\langle S^* \rangle = 1$ by fixing the coefficient of the first Legendre polynomial $P_0$ equal to 1. 

\section{Inviscid results}
\label{sec:res}
 In the Introduction, we asked how distributed flexibility modifies propulsion in comparison to uniform flexibility. Before presenting our results on the kinematics and propulsive characteristics of flapping plates with distributed flexibility, we will briefly review the results for uniform flexibility from \citet{floryan2018clarifying} in order to contextualize our results. All of our results for uniformly flexible plates will be presented relative to rigid plates. For example, we will present the mean thrust that a uniformly flexible plate produces relative to the mean thrust that an otherwise identical rigid plate produces.

\subsection{Propulsive characteristics of flapping plates with uniform flexibility}
\label{sec:uni}
\begin{figure}
  \begin{center}
  \includegraphics[width=\linewidth]{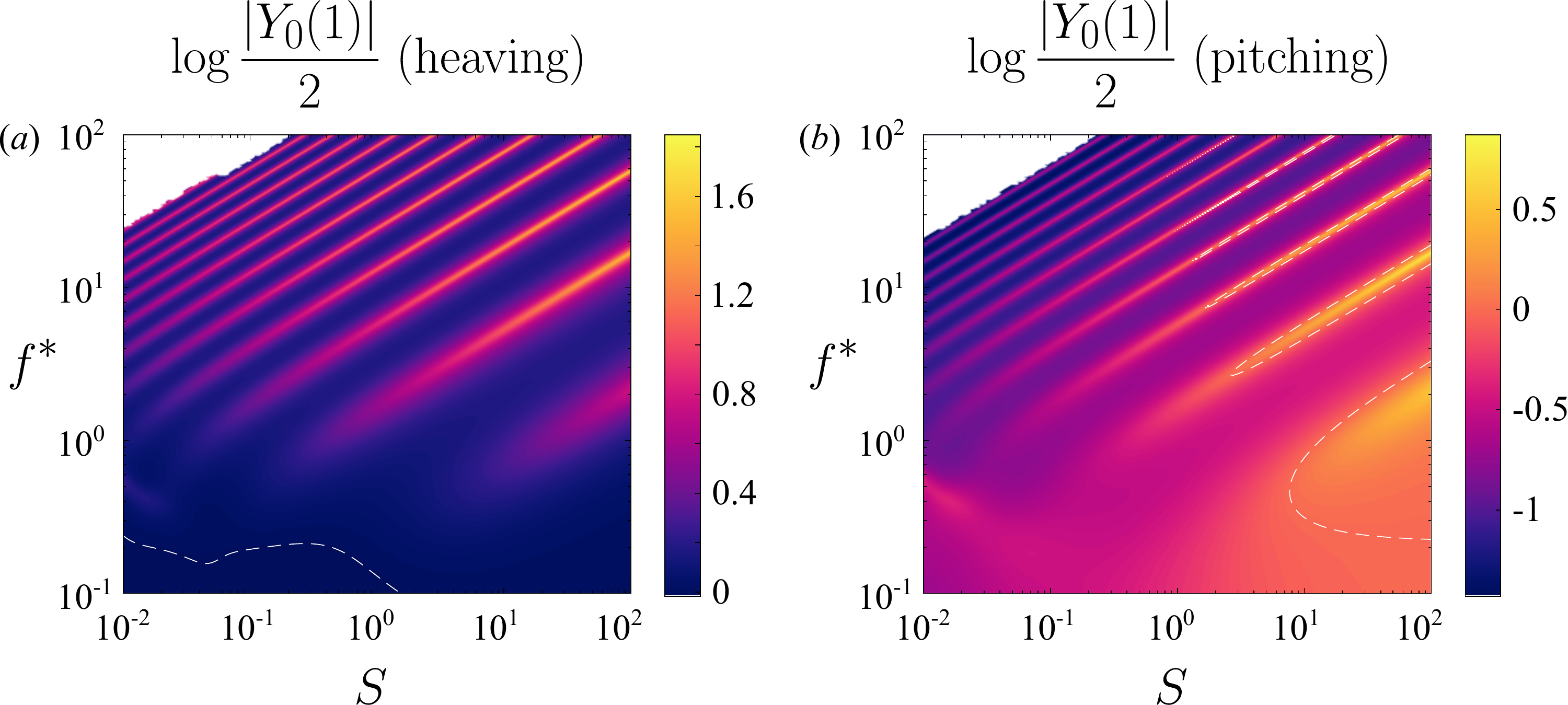}
  \end{center}
  \caption{Trailing edge amplitude as a function of reduced frequency $f^*$ and stiffness ratio $S$ for a (a) heaving and (b) pitching plate with $R \equiv 0.01$ relative to that of an equivalent rigid plate. Dashed white lines indicate where the flexible plate has the same trailing edge amplitude as the equivalent rigid plate. Under-resolved areas have been whited out. Results are for a uniformly flexible plate. The mean thrust and mean power coefficients are qualitatively the same as the trailing edge amplitude. }
  \label{fig:teamp_uni}
\end{figure}

The amplitude of the trailing edge deflection, the mean thrust coefficient, and
the mean power coefficient all exhibit the same qualitative behaviour in the
frequency-stiffness plane. For reference, we have plotted the trailing edge
amplitude in figure~\ref{fig:teamp_uni}. For mid-to-high values of the reduced
frequency and stiffness ratio, ridges of local maxima are apparent. These ridges
coincide with the natural frequencies (imaginary parts of the eigenvalues) of
the system, indicating a resonant response. In this region of the
frequency-stiffness plane, the natural frequencies are well approximated by the
quiescent natural frequencies, which are calculated in the limit where the
bending velocity is large compared to the fluid velocity; we provide more
details in Appendix~\ref{sec:quiescent}. The eigenvalues are lightly damped
(small angle relative to the imaginary axis) and well separated, leading to the
sharp ridges observed. The natural frequencies increase as the stiffness ratio
increases, conforming to our intuition based on a clamped Euler-Bernoulli beam
in vacuo, and can be shown to vary as $f^* \sim S^{1/2}$ in this region of the
frequency-stiffness plane. We will refer to these eigenvalues and the
corresponding eigenfunctions as Euler-Bernoulli modes.

The behaviour is quite different when the reduced frequency and stiffness ratio
are low, however. In this region of the frequency-stiffness plane, the resonant
peaks broaden and smear together as the stiffness ratio decreases because the
eigenvalues become more damped and move closer to each other. A ridge aligned in
the direction opposite to the other ridges emerges, with the frequency
\emph{decreasing} as stiffness increases, although we note that the mean thrust and mean
power for a pitching plate actually become negative here, in contrast to the
trailing edge amplitude for a pitching plate. Whereas the previous ridges
coincided with the natural frequencies of the Euler-Bernoulli modes, this ridge
aligns with the natural frequencies of a `flutter mode', a mode that becomes
unstable for low enough stiffness ratio and induces flutter in the beam (as seen
in a flag flapping in the wind, for example) \citep{alben2008flapping}.
With decreasing stiffness ratio,
Euler-Bernoulli modes are essentially replaced by flutter modes, with the
replacement occurring at lower values of the stiffness ratio for higher-order
modes. The flutter modes are weakly damped compared to the Euler-Bernoulli
modes, leading to ridges aligned with the flutter modes.

\begin{figure}
  \begin{center}
  \includegraphics[width=\linewidth]{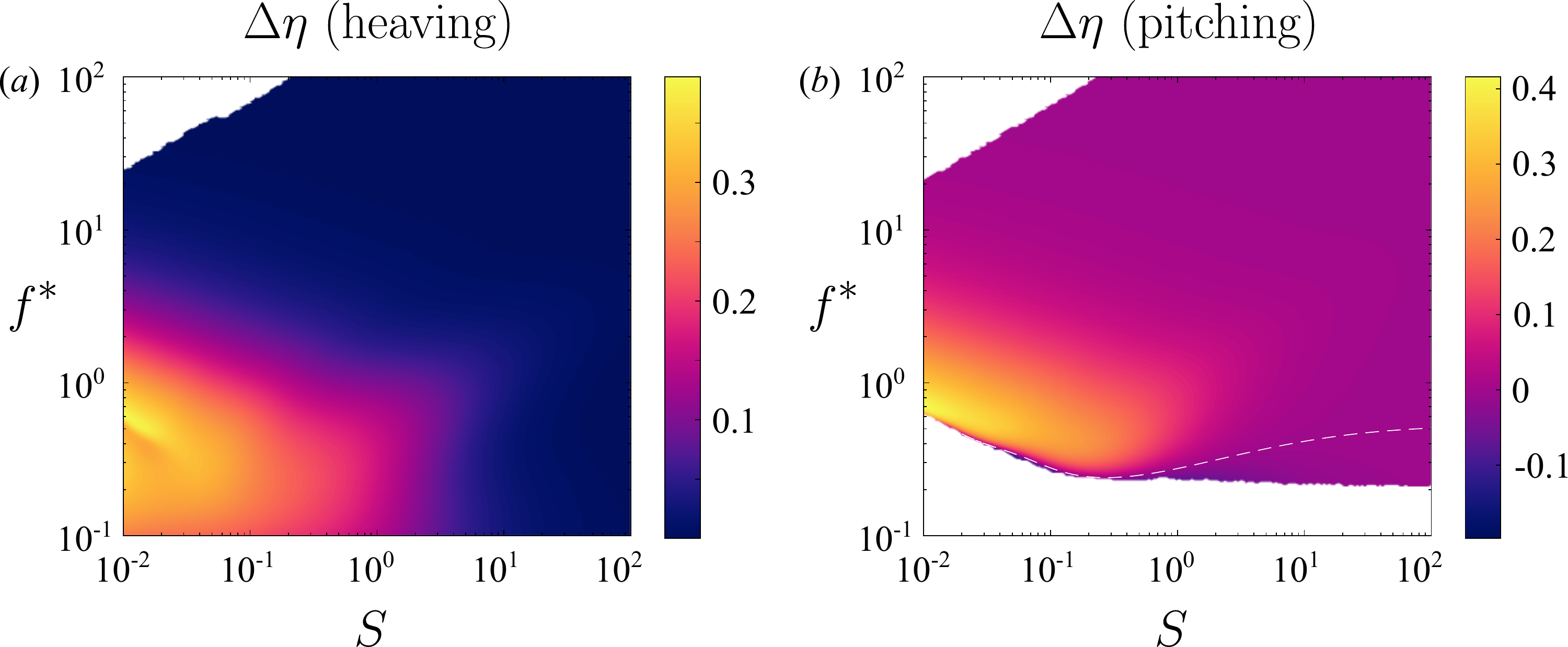}
  \end{center}
  \caption{Efficiency as a function of reduced frequency $f^*$ and stiffness ratio $S$ for a (a) heaving and (b) pitching plate with $R \equiv 0.01$ relative to that of an equivalent rigid plate. Dashed white lines indicate where the flexible plate has the same efficiency as the equivalent rigid plate. Under-resolved areas and areas that produce negative efficiency have been whited out. Results are for a uniformly flexible plate. }
  \label{fig:etadiff_uni}
\end{figure}

The efficiency behaves very differently from the trailing edge amplitude, mean thrust, and mean power. In figure~\ref{fig:etadiff_uni}, we have plotted the difference in efficiency between a uniformly flexible plate and an otherwise identical rigid plate. Whereas the trailing edge amplitude, mean thrust, and mean power have ridges of local maxima aligned with the natural frequencies, the efficiency has a single broad region of high values in the frequency-stiffness plane. Elsewhere in the plane, the local maxima in thrust and efficiency cancel each other exactly, resulting in flat efficiency. The region of high efficiency is aligned with the natural frequencies of the flutter mode. The flutter mode induces travelling wave kinematics in the plate, which is known to be highly efficient \citep{wu1961swimming}. It is worth keeping in mind that increases in efficiency are often accompanied by decreases in thrust, as is evinced by the efficiency plot for a pitching flexible plate (figure~\ref{fig:etadiff_uni}b). For the pitching plate, the cutoff where efficiency and thrust become negative is aligned with the natural frequencies of the flutter mode; although the flutter mode induces efficient kinematics, the kinematics lead to low thrust. We must be wary of low values of thrust when drag is present in the system. 

\subsection{Propulsive characteristics of flapping plates with distributed flexibility}
\label{sec:dis}
\begin{figure}
  \begin{center}
  \includegraphics[width=\linewidth]{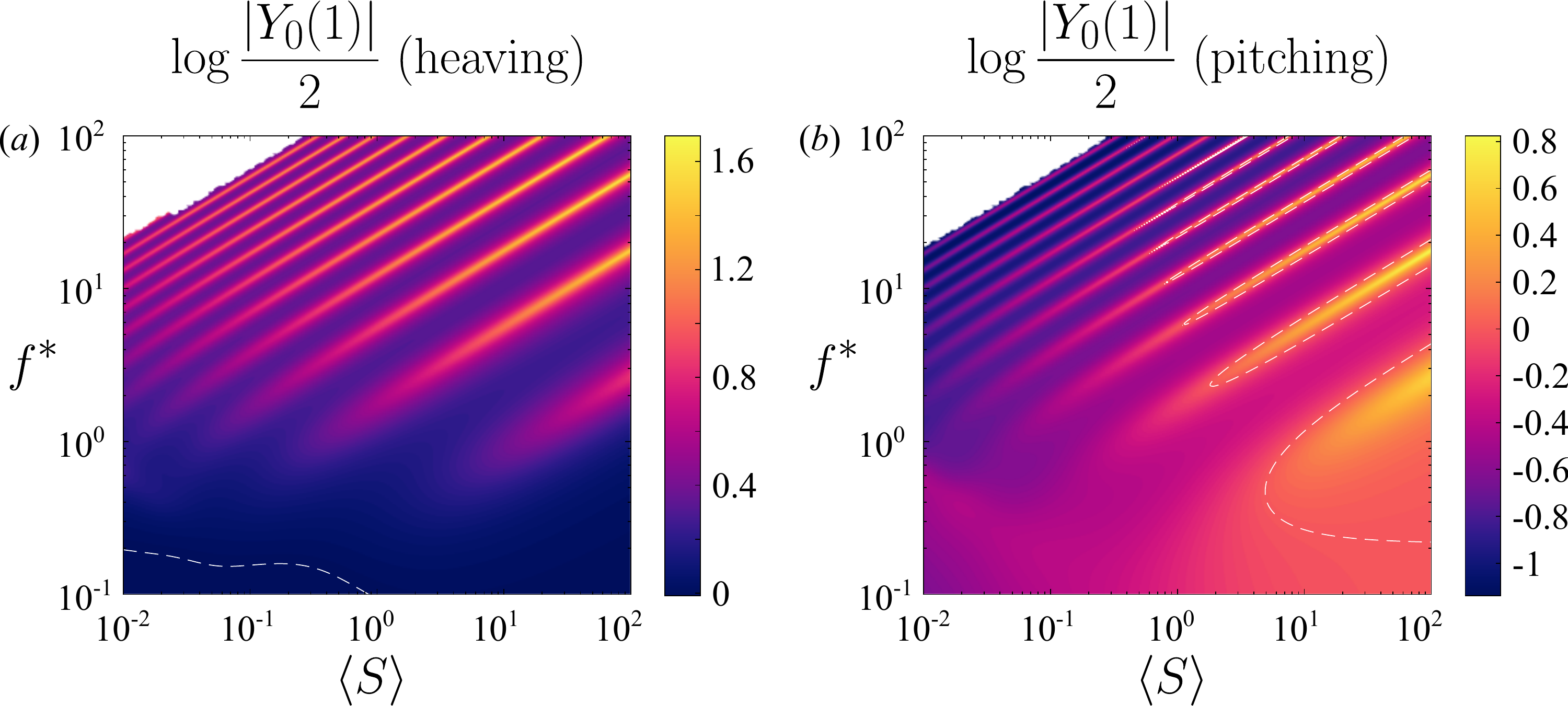}
  \end{center}
  \caption{Analog of figure~\ref{fig:teamp_uni}, for a stiffness distribution $S^*(x) = 1 - 0.9x$ (stiff leading edge). }
  \label{fig:teamp_LEstiff}
\end{figure}

\begin{figure}
  \begin{center}
  \includegraphics[width=\linewidth]{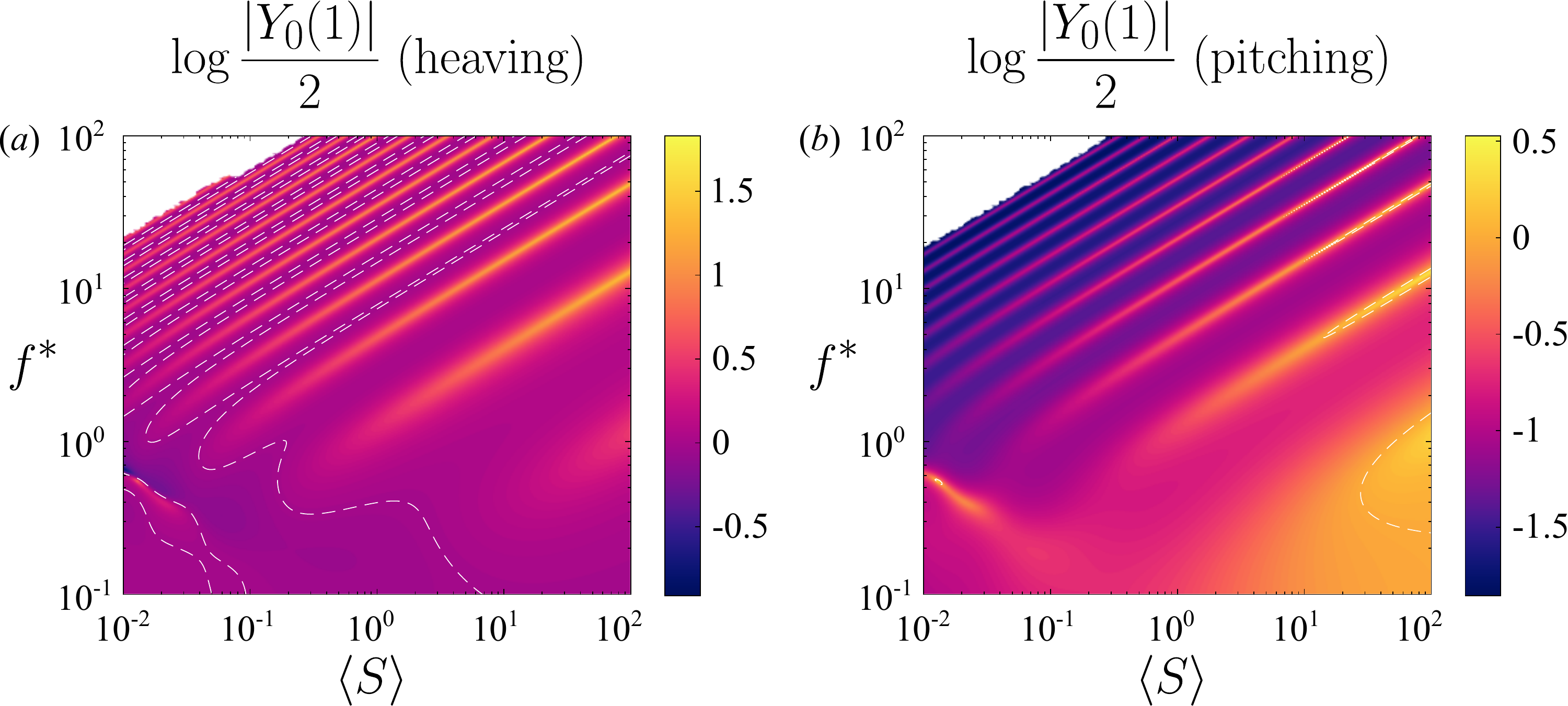}
  \end{center}
  \caption{Analog of figure~\ref{fig:teamp_uni}, for a stiffness
    distribution $S^*(x) = 1 + 0.9x$ (soft leading edge).}
  \label{fig:teamp_LEsoft}
\end{figure}

We begin by considering linear stiffness distributions, which are described by a single parameter $\text{d}S^*/\text{d}x$. Qualitatively, flexible plates with linearly distributed stiffness are the same as flexible plates with uniformly distributed stiffness. In figures~\ref{fig:teamp_LEstiff} and~\ref{fig:teamp_LEsoft}, we have plotted the trailing edge amplitude of plates with a stiff leading edge ($\text{d}S^*/\text{d}x = -0.9$) and a soft leading edge ($\text{d}S^*/\text{d}x = 0.9$), respectively. As before, the trailing edge amplitude is qualitatively representative of the mean thrust and power coefficients. The plates with stiff and soft leading edges show the same trends as a uniformly flexible plate: sharp resonant ridges for high reduced frequencies and stiffness ratios; broadening and smearing of the ridges for low reduced frequencies and stiffness ratios; and emergence of flutter modes for low stiffness ratios. 

\begin{figure}
  \begin{center}
  \includegraphics[width=\linewidth]{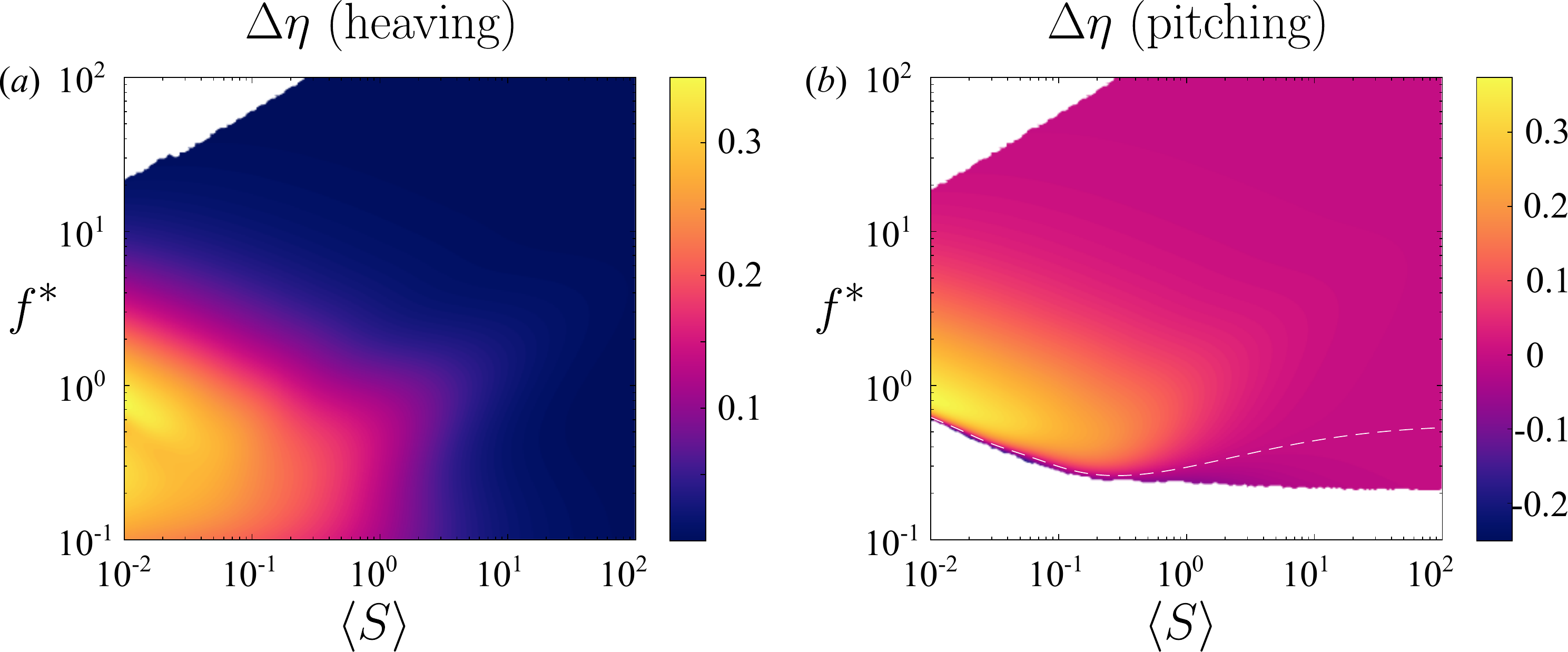}
  \end{center}
  \caption{Analog of figure~\ref{fig:etadiff_uni}, for a stiffness distribution $S^*(x) = 1 - 0.9x$ (stiff leading edge).}
  \label{fig:etadiff_LEstiff}
\end{figure}

\begin{figure}
  \begin{center}
  \includegraphics[width=\linewidth]{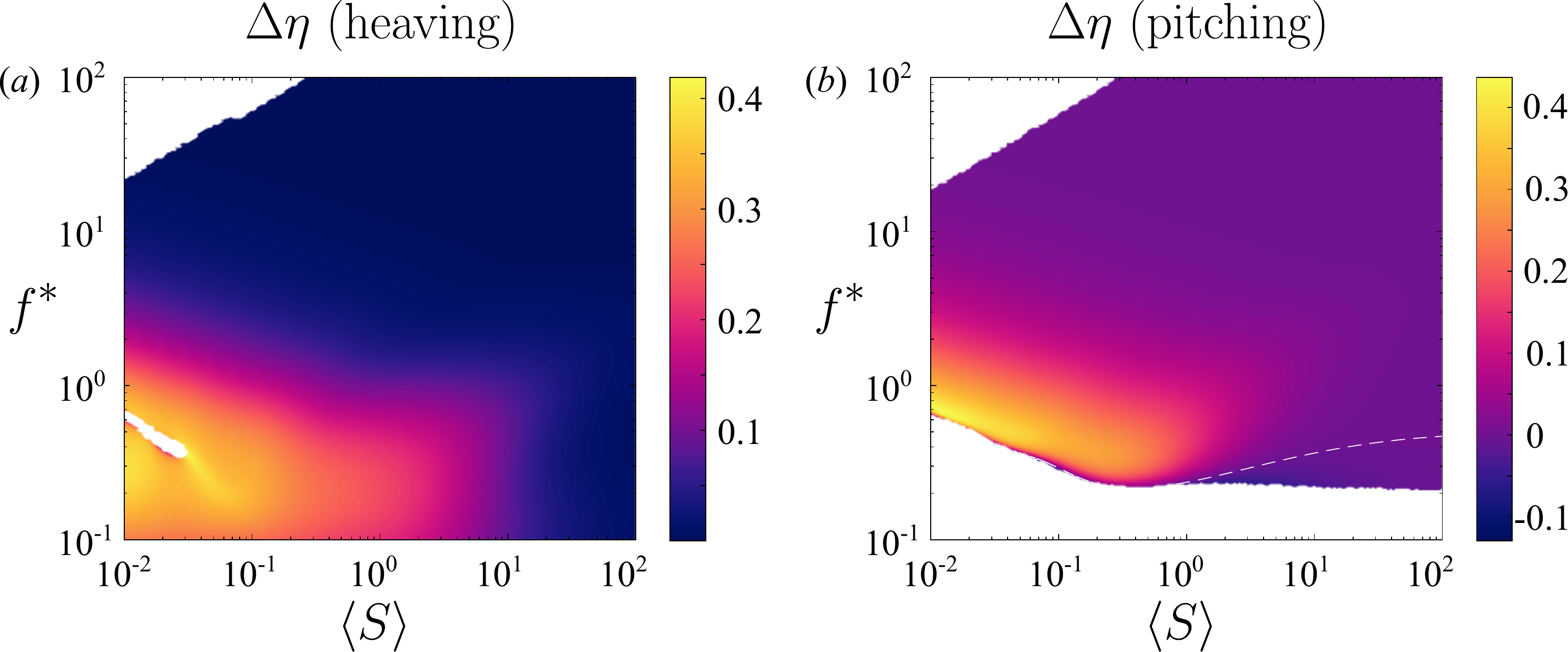}
  \end{center}
  \caption{Analog of figure~\ref{fig:etadiff_uni}, for a stiffness distribution $S^*(x) = 1 + 0.9x$ (soft leading edge).}
  \label{fig:etadiff_LEsoft}
\end{figure}

The behaviour of the efficiency does not change either. In figures~\ref{fig:etadiff_LEstiff} and~\ref{fig:etadiff_LEsoft}, we have plotted the difference in efficiency between plates with stiff and soft leading edges, respectively, and a rigid plate. In both cases, the efficiency does not have any resonant ridges, but does have a broad region of high values for low reduced frequencies and stiffness ratios. Just as for the trailing edge amplitude, mean thrust coefficient, and mean power coefficient, the efficiency of plates with distributed flexibility follows the same trends and for uniformly flexible plates. 

By and large, there are no qualitative differences between plates with uniform stiffness and plates with linearly distributed stiffness. The behaviour of the measures of propulsive performance is dominated by the eigenvalues of the system, which are qualitatively the same for different distributions of stiffness: Euler-Bernoulli modes govern the performance when the stiffness ratio is high, whereas flutter modes emerge and govern the performance when the stiffness ratio is low. We posit that the behaviour does not change qualitatively for higher-order distributions of stiffness. Although there are no qualitative differences, there may be important quantitative differences, and we shall explore them in the next section.

\section{Optimal stiffness distributions}
\label{sec:opt}
As discussed in the Introduction, the literature has shown that varying the
distribution of stiffness quantitatively changes the propulsive performance of
flexible plates. In some cases \citep{moore2015torsional,
  kancharala2016optimal}, the stiffness distribution was optimized in order to
achieve the greatest thrust/speed or greatest efficiency/lowest cost of
transport. Optimal stiffness distributions differed qualitatively for different
mass ratios. For a low mass ratio, relevant to swimmers, concentrating the
stiffness towards the leading edge maximized thrust and efficiency
\citep{kancharala2016optimal}, but the authors did not control for mean
stiffness and only studied a few frequencies. Here, we will calculate optimal
stiffness distributions at every point in the frequency-stiffness plane we have
explored. In particular, for every combination of reduced frequency and mean
stiffness, we solve for the distribution of stiffness that: (a) maximizes
thrust; (b) minimizes power; and (c) maximizes efficiency.

\begin{figure}
  \begin{center}
  \includegraphics[width=\linewidth]{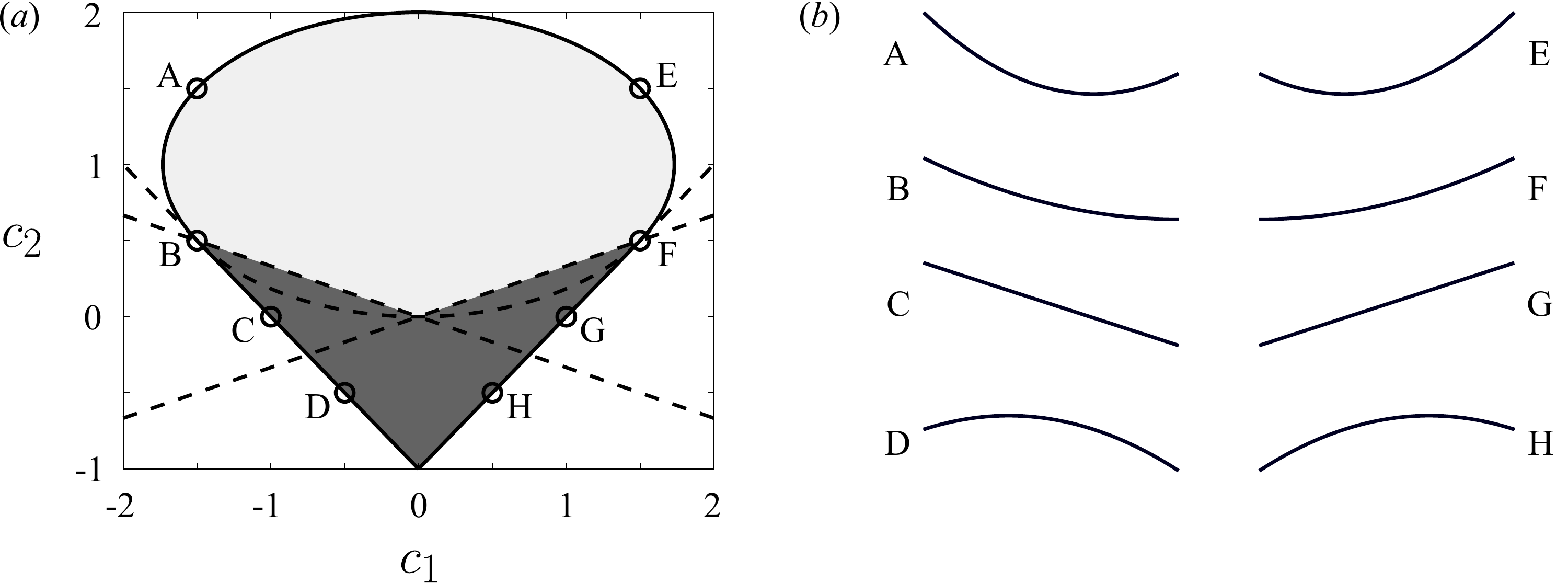}
  \end{center}
  \caption{(a) Feasible set of quadratic flexibility distributions.The dark region
   contains stiffness distributions whose
    minima are at the leading/trailing edge, and the light region contains
    stiffness distributions whose minima are at an interior point of the
    plate. We have drawn some representative distributions in (b), corresponding
    to the circles in (a).}
  \label{fig:feas}
\end{figure}

For now, we will limit ourselves to quadratic distributions of stiffness, but we will end with how we expect our results to generalize to higher-order distributions. The distribution of stiffness can be written as
\begin{equation}
  \label{eq:opt1}
  S^* = P_0 + c_1 P_1 + c_2 P_2,
\end{equation}
where $P_i$ are the Legendre polynomials (written out in~\eqref{eq:param2}), and $c_i$ are the parameters we optimize over. The coefficient multiplying $P_0$ is fixed to 1 so that $\langle S^* \rangle = 1$. Furthermore, we must restrict $c_1$ and $c_2$ so that the stiffness is non-negative on the plate. The physical constraint of non-negativity leads to
\begin{equation}
  \label{eq:opt2}
  \begin{cases}
  -3c_2^2 + 6c_2 - c_1^2 \ge 0 &\text{if $-3c_2 \le c_1 \le 3c_2$}\\
  1 \pm c_1 + c_2 \ge 0 &\text{otherwise.}
  \end{cases}
\end{equation}
The feasible set is drawn in figure~\ref{fig:feas}, along with some representative stiffness distributions. The dark region contains stiffness distributions whose minima are at the leading/trailing edge, and the light region contains stiffness distributions whose minima are at an interior point of the plate.

Altogether, we have a nonlinear constrained optimization problem, with both the
objective function and the constraints being nonlinear in the optimization
variables. Since the feasible set is the union of an ellipse and a convex
polyhedron, the original optimization problem can be split into two optimization
problems with linear constraints (the ellipse can be described by linear
constraints in polar coordinates). We solve the optimization problem using
MATLAB's default interior-point algorithm \citep{matlab2016a}. The objective
functions used here are non-convex, so we use many initial guesses to be 
confident that we have found a global optimum; this is feasible only because of
the speed of the numerical method. 

\subsection{Linear stiffness distributions}
\label{sec:lin}
\begin{figure}
  \begin{center}
  \includegraphics[width=\linewidth]{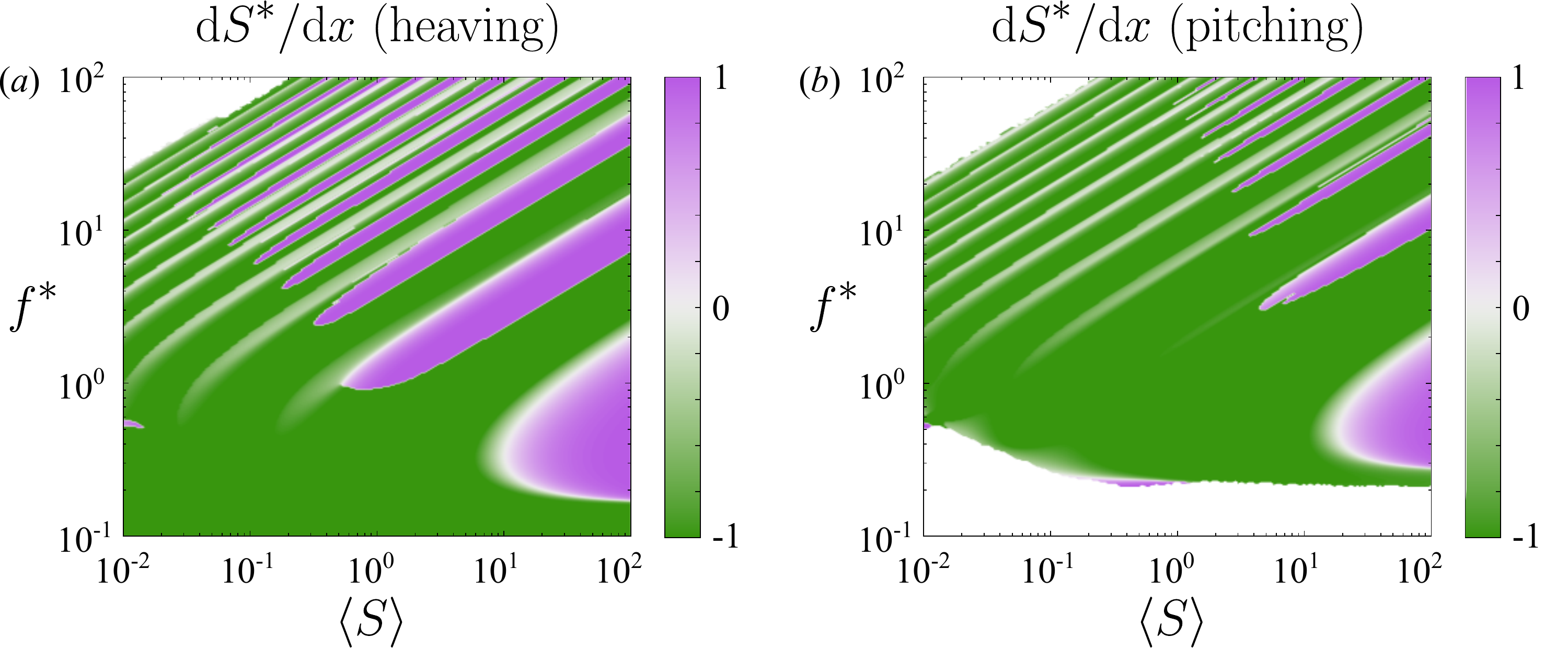}
  \end{center}
  \caption{Thrust-maximizing linear stiffness distribution as a function of reduced frequency $f^*$ and mean stiffness ratio $\langle S \rangle$ for a (a) heaving and (b) pitching plate with $R \equiv 0.01$. Under-resolved areas and areas that produce negative thrust have been whited out. }
  \label{fig:thrust_opt_distr}
\end{figure}

\begin{figure}
  \begin{center}
  \includegraphics[width=\linewidth]{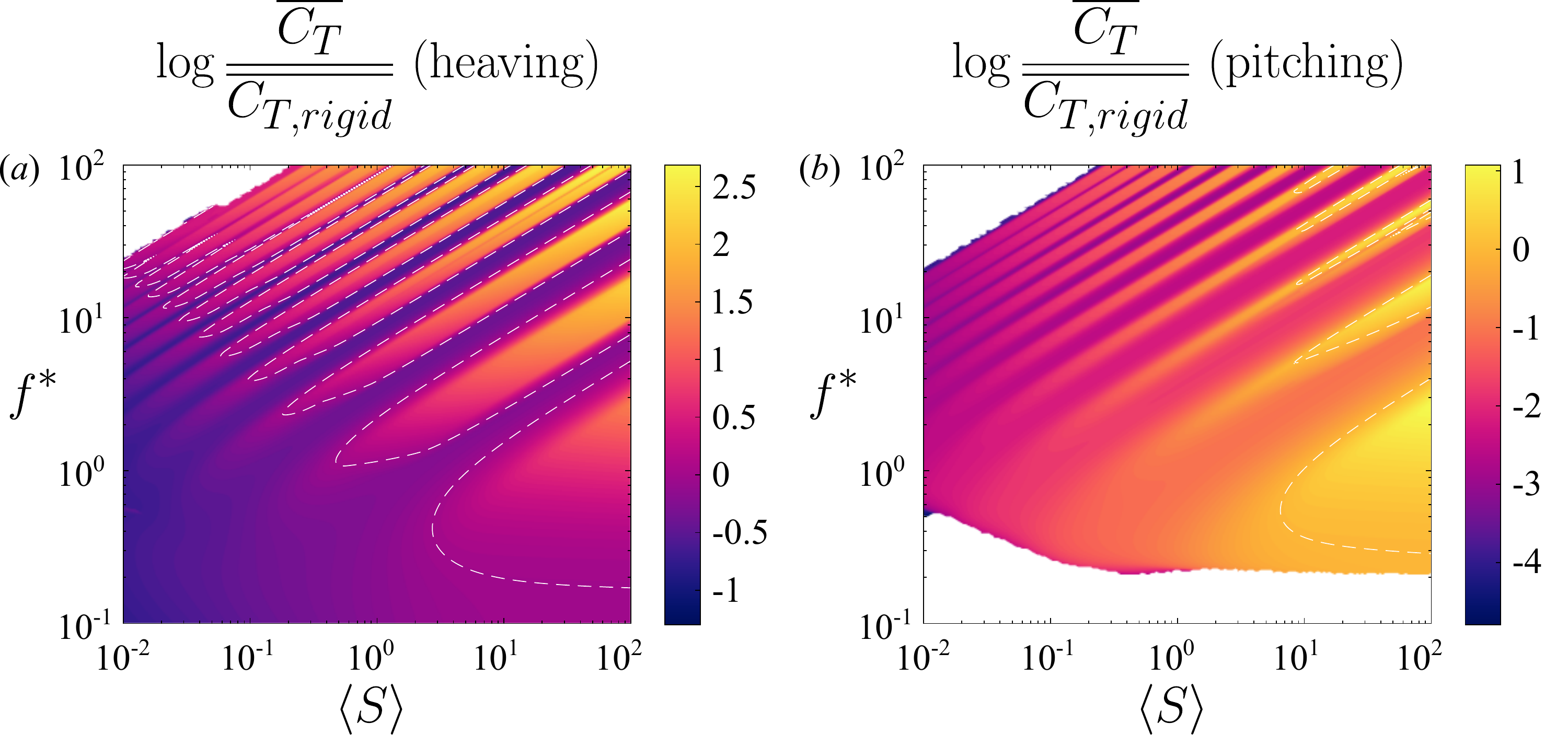}
  \end{center}
  \caption{Thrust coefficient of a plate with the stiffness distribution shown in figure~\ref{fig:thrust_opt_distr} relative to that of an equivalent rigid plate. Dashed white lines indicate where the flexible plate has the same thrust coefficient as the equivalent rigid plate. }
  \label{fig:thrust_opt}
\end{figure}

We begin by calculating optimal linear stiffness distributions, in which case the only optimization parameter is the slope of the stiffness distribution, $\text{d}S^*/\text{d}x$. When $\text{d}S^*/\text{d}x < 0$, we say that the plate has a stiff leading edge, and when $\text{d}S^*/\text{d}x > 0$, we say that the plate has a soft leading edge. In figure~\ref{fig:thrust_opt_distr}, we have plotted the optimal (thrust-maximizing) linear stiffness distribution, with the attendant optimal mean thrust coefficient plotted in figure~\ref{fig:thrust_opt}. There is a clear distinction in behaviour between high-stiffness regions (where Euler-Bernoulli modes dominate the behaviour) and low-stiffness regions (where flutter modes dominate), so we will discuss them in turn. 

When the Euler-Bernoulli modes dominate the response, the optimal stiffness distribution at a given reduced frequency and mean stiffness ratio is the one that has a natural frequency at that frequency of actuation. This is consistent with our understanding of uniformly stiff plates, where actuating at a natural frequency produces a local maximum in thrust. By tuning the stiffness distribution appropriately, we can tune the natural frequencies of the plate so that they coincide with the frequency of actuation. The ability to tune the locations of natural frequencies broadens the resonant response, thereby broadening the regions of high thrust, as evinced by figure~\ref{fig:thrust_opt}. These results starkly contrast those for a plate with a fixed stiffness distribution, where the resonant response is quite narrow (cf.\ figures~\ref{fig:teamp_uni}, \ref{fig:teamp_LEstiff}, and \ref{fig:teamp_LEsoft}). 

\begin{figure}
  \begin{center}
  \includegraphics[width=\linewidth]{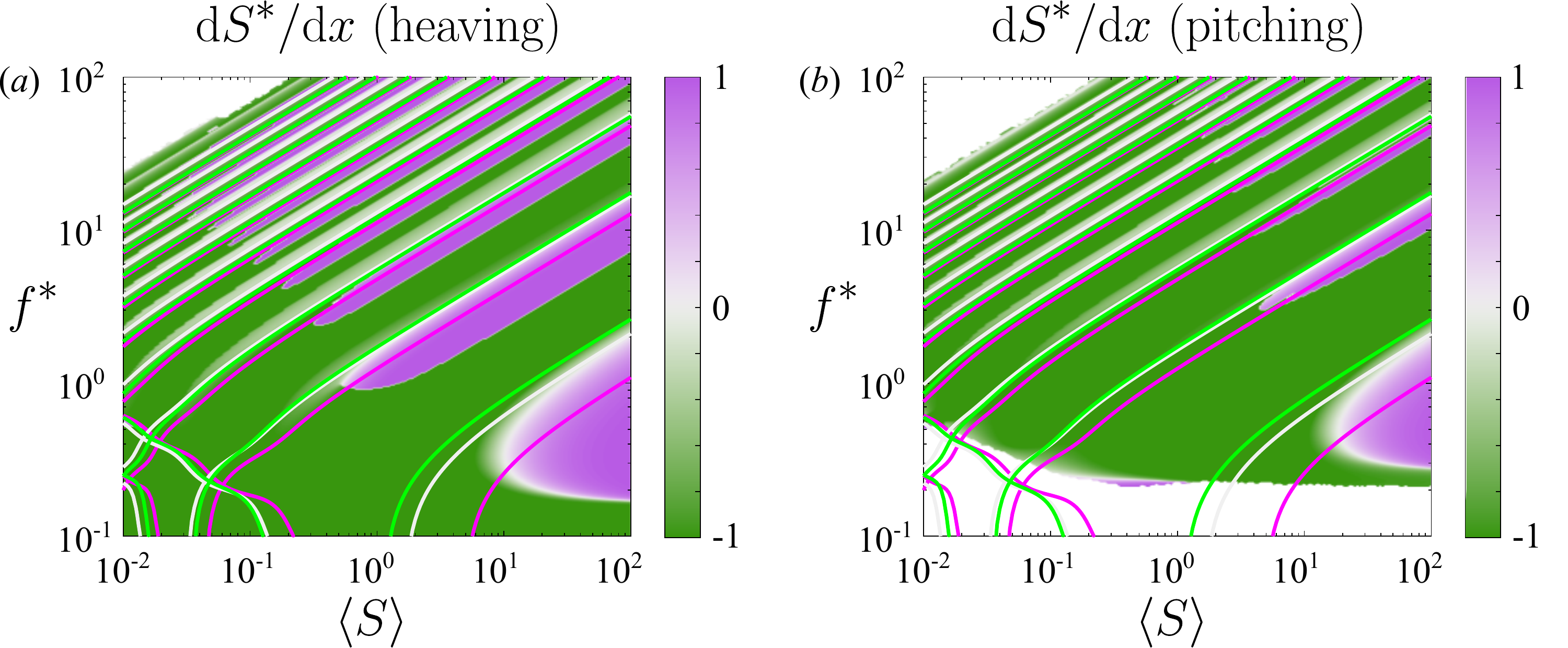}
  \end{center}
  \caption{Same as in figure~\ref{fig:thrust_opt_distr}, but with natural frequencies overlaid as curves for $\text{d}S^*/\text{d}x = -0.9$ (stiff leading edge, green), $\text{d}S^*/\text{d}x = 0$ (uniformly stiff, white), and $\text{d}S^*/\text{d}x = 0.9$ (soft leading edge, purple). }
  \label{fig:thrust_opt_distr_eig}
\end{figure}

A resonant response is not always possible, however. Although being able to modify the stiffness distribution greatly broadens the resonant ridges, there are still valleys of relatively low thrust in between the resonant ridges. This is because natural frequencies of lower-order modes do not overlap with natural frequencies of higher-order modes. To make the situation clear, we have re-plotted the optimal linear stiffness distribution in figure~\ref{fig:thrust_opt_distr_eig} with the natural frequencies for stiff leading edge ($\text{d}S^*/\text{d}x = -0.9$, green), uniformly stiff ($\text{d}S^*/\text{d}x = 0$, white), and soft leading edge ($\text{d}S^*/\text{d}x = 0.9$, purple) plates overlaid as three sets of curves. Clearly, the natural frequency of the first Euler-Bernoulli mode for a plate with a stiff leading edge is nowhere close to the natural frequency of the second Euler-Bernoulli mode for a plate with a soft leading edge, and so on for higher-order modes. The gap between natural frequencies that are attainable with a linear stiffness distribution leads to the valleys in optimal thrust between resonant ridges. 

\begin{figure}
  \begin{center}
  \includegraphics[width=0.5\linewidth]{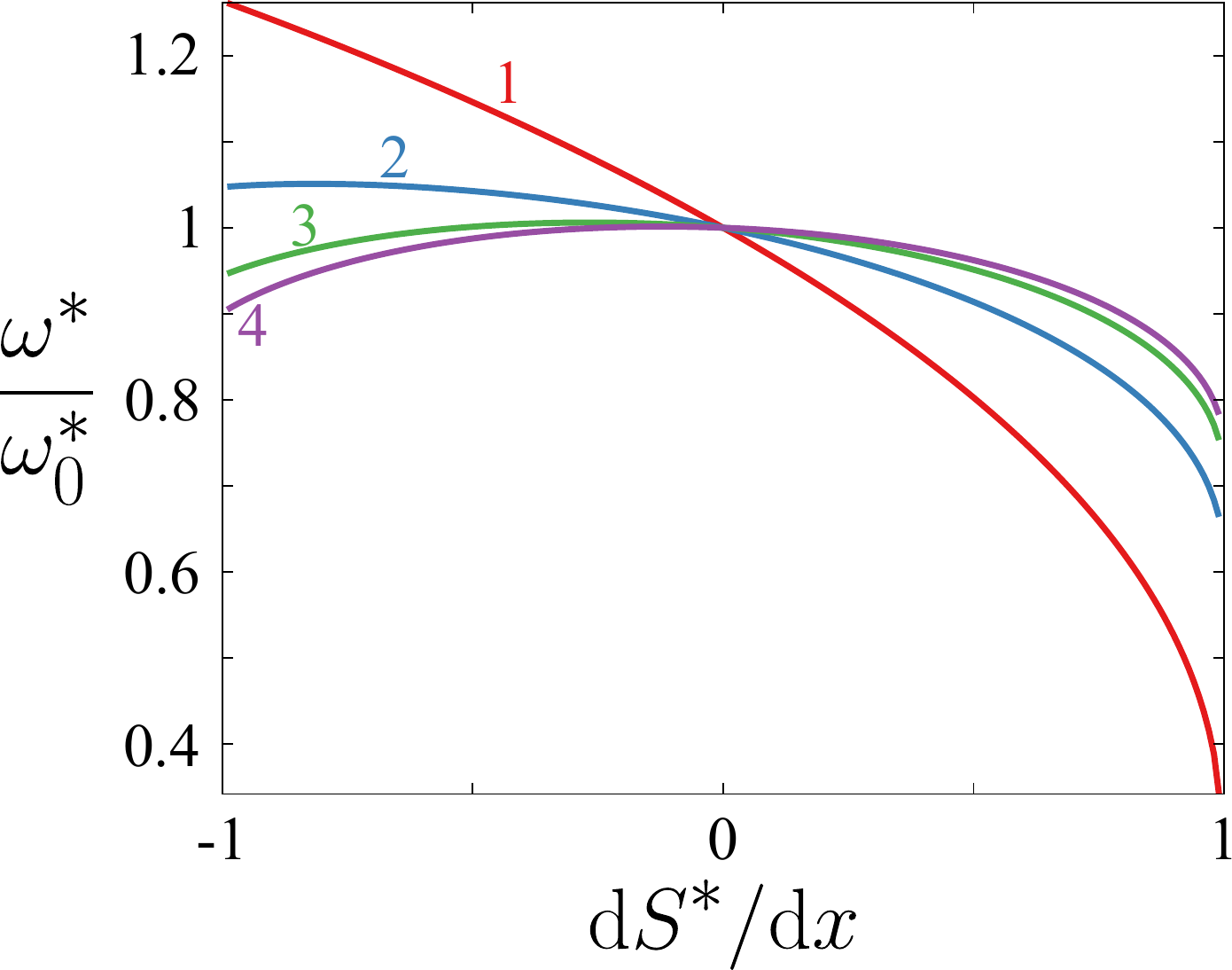}
  \end{center}
  \caption{First four quiescent natural frequencies as a function of the stiffness distribution. The natural frequencies have been normalized by the natural frequencies of a uniformly stiff plate, $\text{d}S^*/\text{d}x = 0$. }
  \label{fig:eigqui}
\end{figure}

Moreover, for high-order modes the natural frequencies do not follow the pattern
we might expect. For the first mode, a plate with a soft leading edge has the
lowest natural frequency, and a plate with a stiff leading edge has the highest
natural frequency, as one might expect. By the third mode, however, a uniformly
stiff plate has a higher natural frequency than plates with stiff or soft
leading edges. This is more clearly shown in figure~\ref{fig:eigqui}, where we
have plotted the quiescent natural frequencies for the first four modes as a
function of the stiffness distribution. (For just this plot, we have
non-dimensionalized time using the bending time scale $t_{bend}$, as explained
in Appendix~\ref{sec:quiescent}, yielding $\omega^* = \omega t_{bend}$, where
$\omega = 2\upi f$ is the dimensional angular frequency.) We have rescaled the
quiescent natural frequencies so that they are plotted relative to the values
for a uniformly stiff plate. For third- and higher-order modes, uniformly stiff
plates have higher natural frequencies than plates with a stiff (or soft)
leading edge. Consequently, the relation between stiffness distribution and
natural frequency is not one-to-one: plates with a stiff leading edge may have
the same natural frequency as plates with a soft leading edge. Because of this, it
is not possible to represent a plate with distributed stiffness as a uniform plate
with some effective stiffness. 

\begin{figure}
  \begin{center}
  \includegraphics[width=\linewidth]{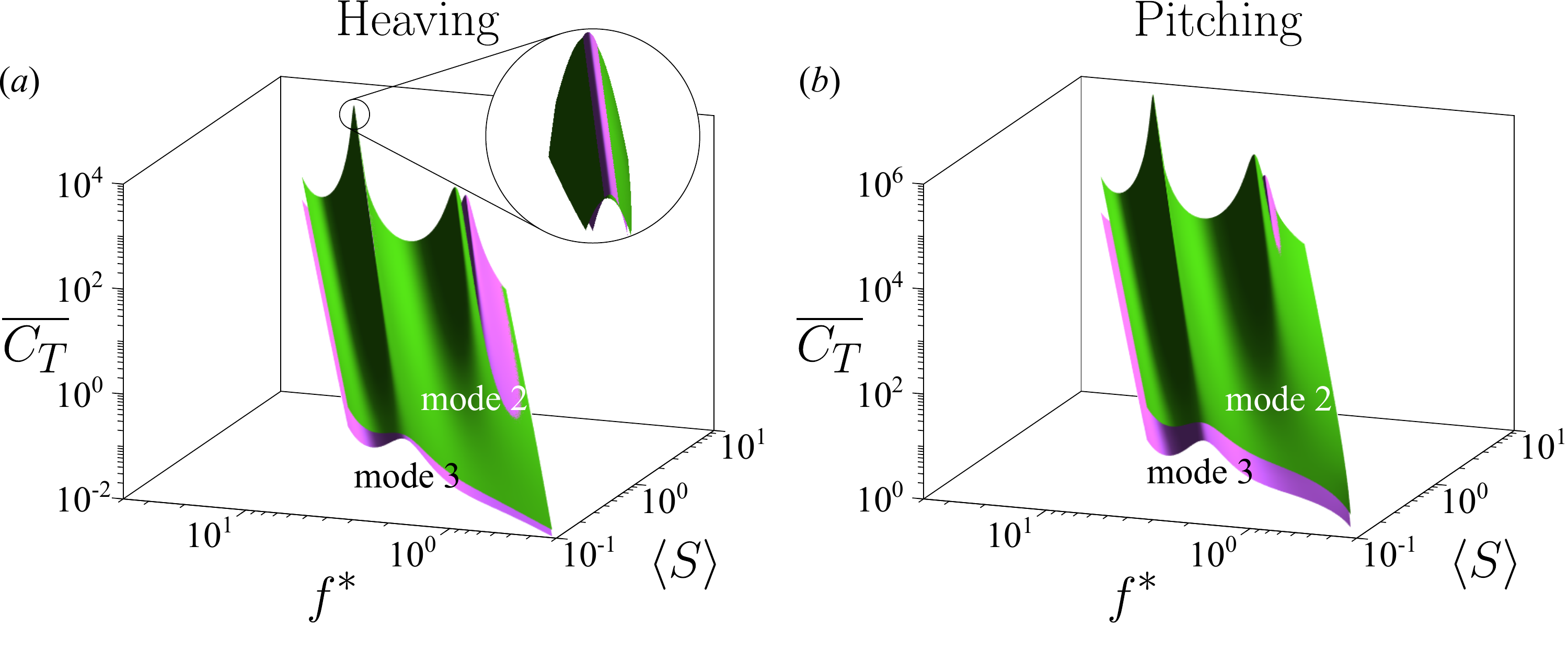}
  \end{center}
  \caption{Thrust coefficient as a function of reduced frequency $f^*$ and mean stiffness ratio $\langle S \rangle$ for a (a) heaving and (b) pitching plate with $R \equiv 0.01$. The green surface corresponds to a plate with a stiff leading edge ($\text{d}S^*/\text{d}x = -0.9$), and the purple surface corresponds to a plate with a soft leading edge and matched third natural frequency ($\text{d}S^*/\text{d}x \in [0.416,0.445]$). }
  \label{fig:thrust_comp}
\end{figure}

When multiple stiffness distributions have the same natural frequency, which distribution is preferred? To provide insight into this question, we have calculated the thrust produced by two plates, one with a stiff leading edge and one with a soft leading edge, with both plates having the same third natural frequency. (The plate with the stiff leading edge has $\text{d}S^*/\text{d}x = -0.9$, while the plate with the soft leading edge has $\text{d}S^*/\text{d}x$ varying from 0.416 to 0.445 over the range of $\langle S \rangle$ considered so that its third natural frequency is the same as that of the plate with the stiff leading edge.) The thrusts produced by the plates are plotted as surfaces in figure~\ref{fig:thrust_comp}, with the green surface corresponding to the plate with a stiff leading edge, and the purple surface corresponding to the plate with a soft leading edge. The surfaces are plotted on top of each other so that the surface that is visible from above has greater thrust. When heaved around the third natural frequency, the plate with a stiff leading edge produces more thrust than the plate with a soft leading edge except for a very tight range of frequencies centered about the natural frequency, as shown in the close-up view. When pitched around the third natural frequency, the plate with a stiff leading edge produces more thrust than the plate with a soft leading edge at all frequencies near the natural frequency. Except when actuated right at the natural frequency, the plate with a stiff leading edge is preferred over the plate with a soft leading edge and equal natural frequency when it comes to thrust production. 

In a similar vein, which stiffness distribution is preferred when the plate is actuated away from a resonant frequency, that is, in the resonant gaps seen in figure~\ref{fig:thrust_opt}? The results show that a plate with a stiff leading edge is always preferred. In the resonant gaps, plates with a stiff leading edge produce the greatest trailing edge amplitude, leading to the greatest thrust production. This is true in the entire region of the frequency-stiffness plane dominated by Euler-Bernoulli-type behaviour: the stiffness distribution that produces the greatest trailing edge amplitude also produces the greatest thrust. 

The last observation we make about optimal thrust production in the region dominated by Euler-Bernoulli behaviour concerns the difference between heaving and pitching plates. For a heaving plate, the regions where a soft leading edge is preferred are more expansive than for a pitching plate. In particular, regions where a very soft leading edge is preferred for a heaving plate are replaced by a stiff leading edge for a pitching plate (these are regions where the frequency of actuation is close but not equal to a natural frequency of a plate with a soft leading edge). To understand why, consider a plate starting at rest with a soft leading edge in the limiting case $\text{d}S^*/\text{d}x = 1$. In this case, the stiffness at the leading edge is zero. When we pitch such a plate at the leading edge, no moment will be generated at the leading edge since the stiffness there is zero. Consequently, there will be no deflection, the plate will remain parallel to the flow, and thus no thrust will be generated. When we heave such a plate, no moment will be generated at the leading edge, but the plate still needs to satisfy the boundary condition at the leading edge. The plate will, therefore, take on something of a sideways L shape, so it will be at an angle to the flow near the leading edge. Because the plate is at an angle to the flow, the fluid will apply a force to the plate and cause it to deflect. Consequently, the plate is able to produce thrust. Plates with very soft leading edges are thus better suited to heaving actuation than pitching actuation. 

\begin{figure}
  \begin{center}
  \includegraphics[width=\linewidth]{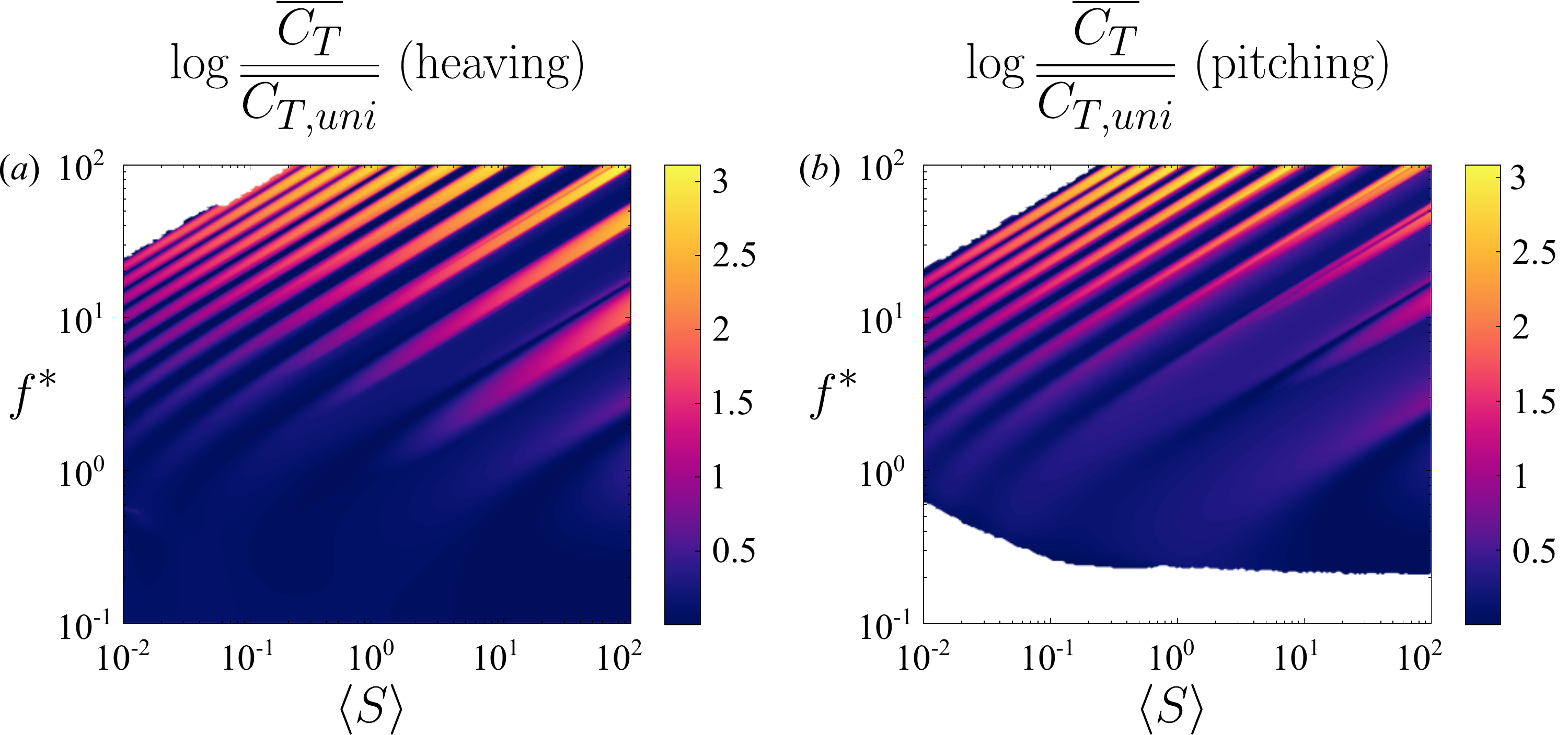}
  \end{center}
  \caption{Analog of figure~\ref{fig:thrust_opt}, but compared to a plate with uniformly distributed stiffness instead of a rigid plate. }
  \label{fig:thrustchangeuni_opt}
\end{figure}

The region dominated by flutter behaviour differs markedly from the region dominated by Euler-Bernoulli behaviour. As we see in figure~\ref{fig:thrust_opt_distr}, a stiff leading edge always produces the most thrust in the flutter region. As we explained in Section~\ref{sec:uni}, in this region the eigenvalues become more damped and move closer to each other, causing the resonant peaks to broaden and smear together. Resonant effects become weak, and the off-resonant behaviour dominates the response. Just as in the region dominated by Euler-Bernoulli behaviour, plates with a stiff leading edge produce the greatest trailing edge amplitude, leading to the greatest thrust production. That being said, the benefits over a plate with uniformly distributed stiffness are modest in this region (the same is true in the resonant gaps), whereas the benefits are quite large when the behaviour is dominated by Euler-Bernoulli modes, which we illustrate in figure~\ref{fig:thrustchangeuni_opt}. 

\begin{figure}
  \begin{center}
  \includegraphics[width=\linewidth]{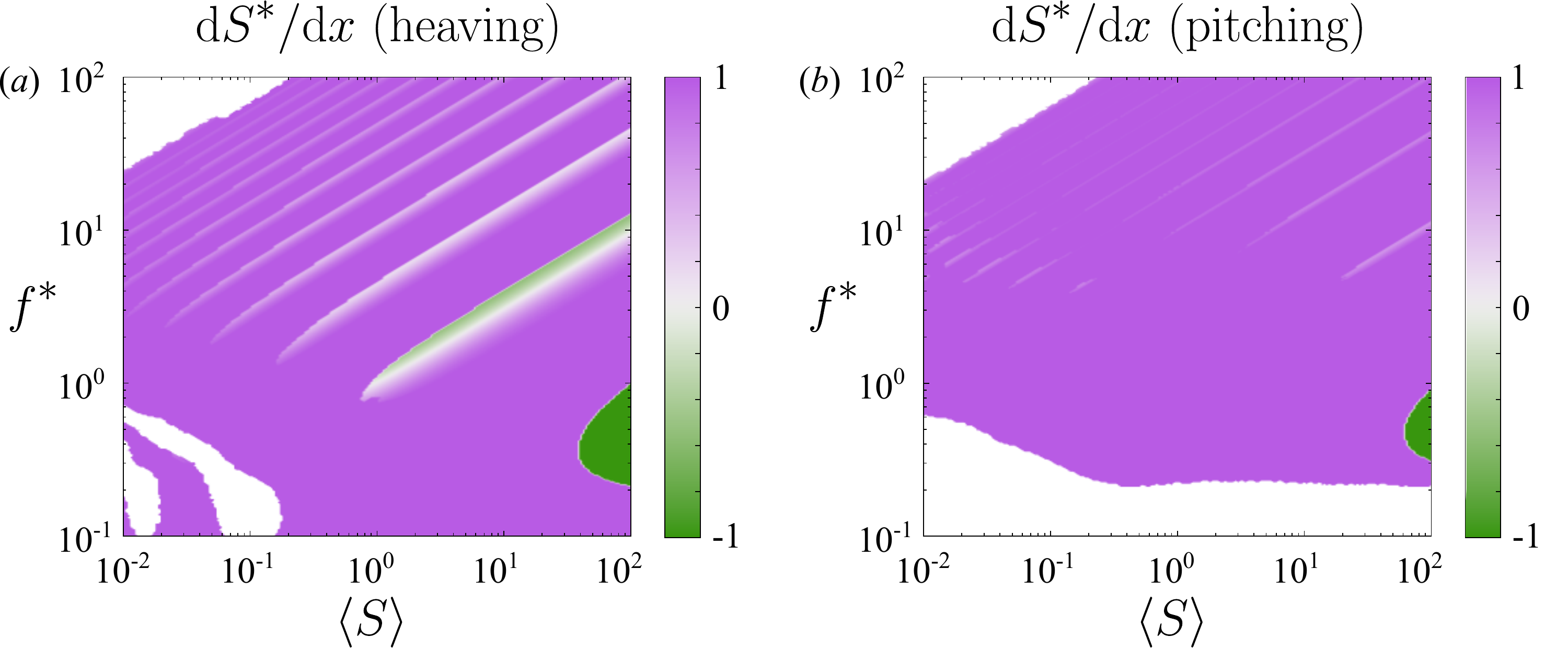}
  \end{center}
  \caption{Analog of figure~\ref{fig:thrust_opt_distr}, but for a linear
    stiffness distribution minimizing power.}
  \label{fig:power_opt_distr}
\end{figure}

\begin{figure}
  \begin{center}
  \includegraphics[width=\linewidth]{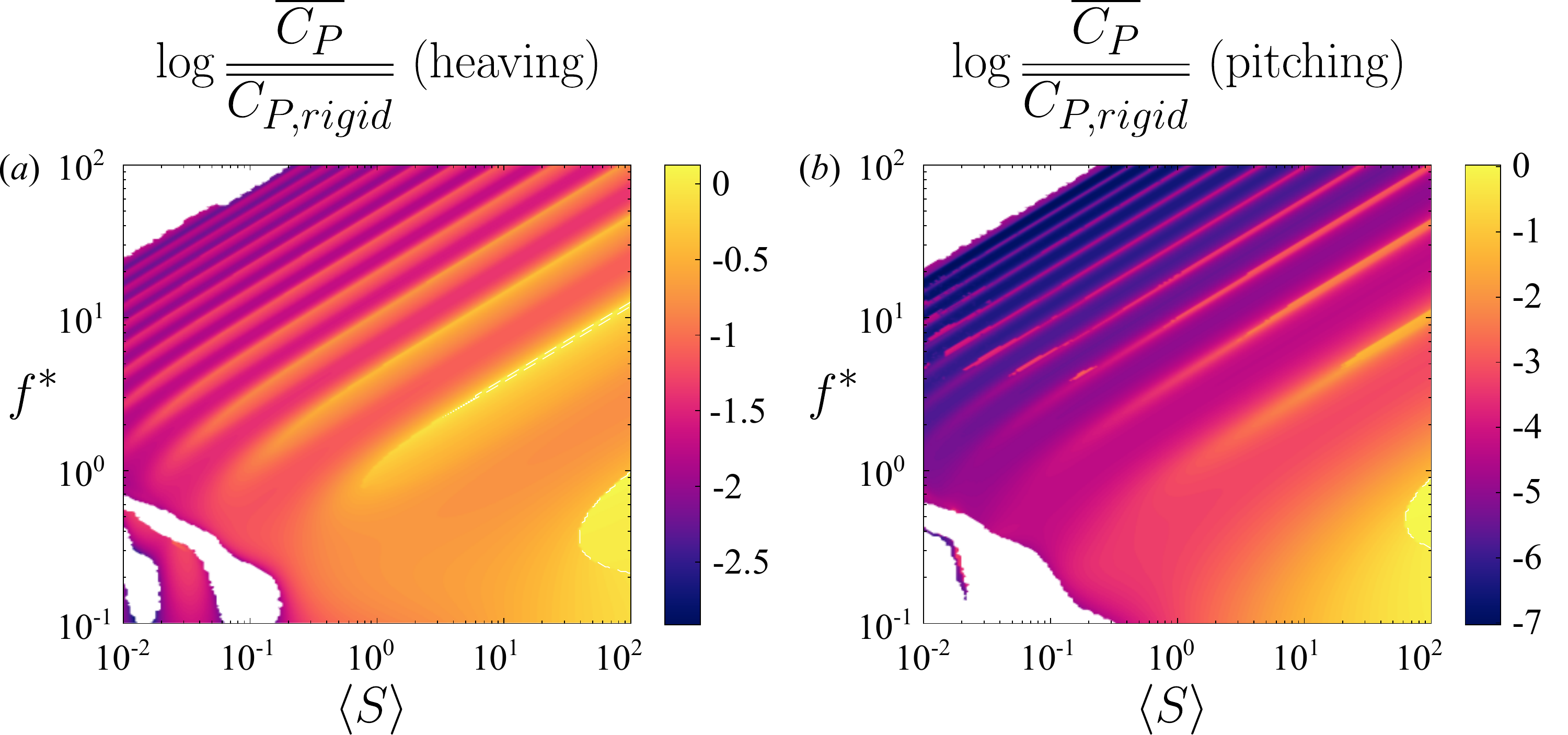}
  \end{center}
  \caption{Power coefficient of a plate with the stiffness distribution shown in figure~\ref{fig:power_opt_distr} relative to that of an equivalent rigid plate. Dashed white lines indicate where the flexible plate has the same power coefficient as the equivalent rigid plate. }
  \label{fig:power_opt}
\end{figure}

We now shift our attention to calculating the linear stiffness distribution that minimizes power consumption. We have plotted the optimal (power-minimizing) linear stiffness distribution in figure~\ref{fig:power_opt_distr}, with the attendant optimal mean power coefficient plotted in figure~\ref{fig:power_opt}. The results are essentially opposite of the results for optimal thrust. In the region where Euler-Bernoulli modes dominate the response, the optimal stiffness distribution is the one whose natural frequencies are far from the frequency of actuation. When maximizing thrust, it was desirable to actuate at resonance, whereas when minimizing power, it is undesirable to actuate at resonance. This is consistent with our understanding of uniformly flexible plates, where actuating at resonance maximizes trailing edge amplitude, thrust, and power. 

In the resonant gaps, where no stiffness distribution has a natural frequency, a soft leading edge is always preferred since it produces a smaller trailing edge amplitude. As we previously explained, a plate with a soft leading edge generally produces a weaker moment at its leading edge, leading to smaller deflection and power consumption. Pitching accentuates this behaviour since the plate is entirely driven by a moment applied at the leading edge, explaining why a soft leading edge occupies a larger area of the frequency-stiffness plane when the plate is pitching than when it is heaving. 

\begin{figure}
  \begin{center}
  \includegraphics[width=\linewidth]{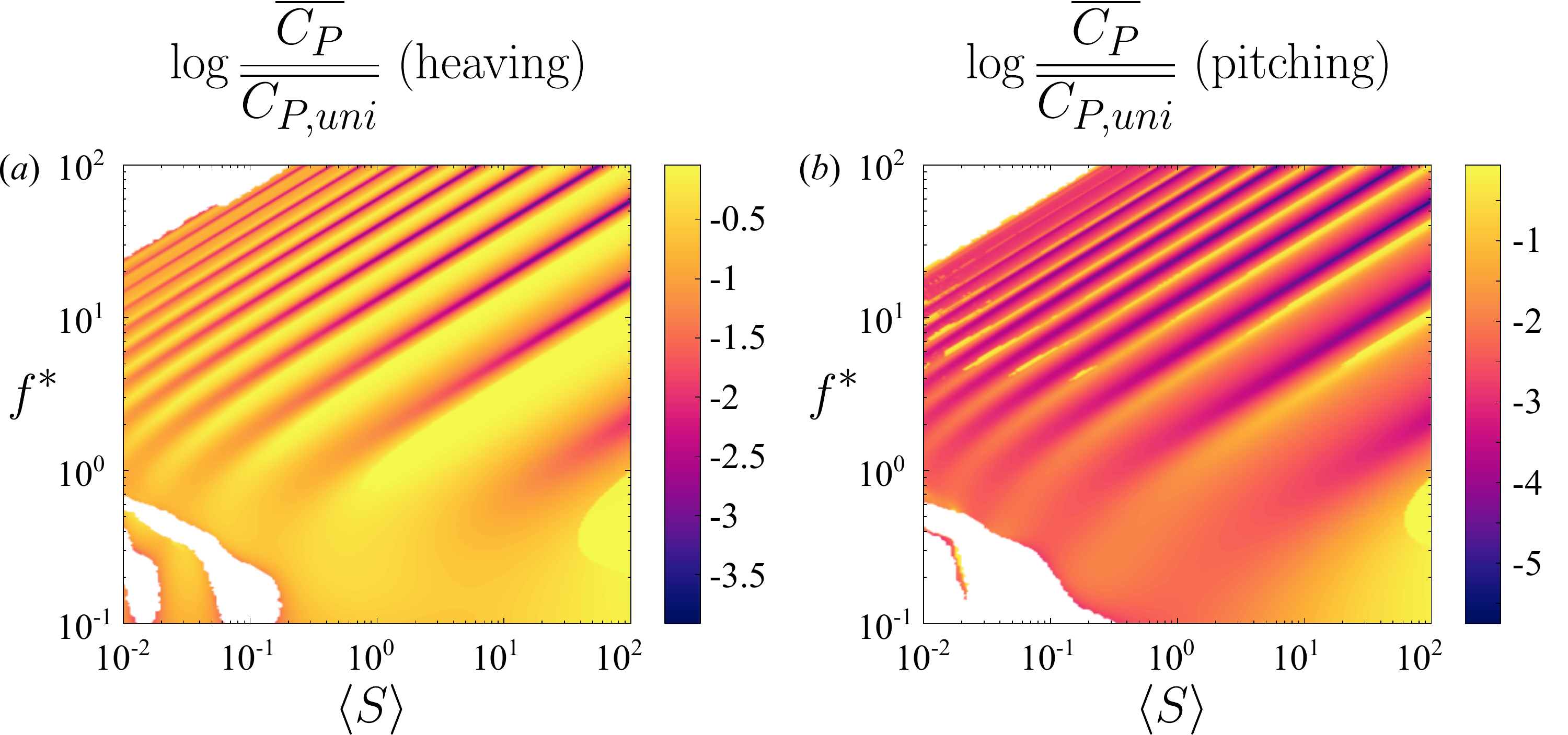}
  \end{center}
  \caption{Analog of figure~\ref{fig:power_opt}, but compared to a plate with uniformly distributed stiffness instead of a rigid plate. }
  \label{fig:powerchangeuni_opt}
\end{figure}

In the region dominated by flutter behaviour, a soft leading edge is also preferred. The effects of resonance are diminished in this region since the eigenvalues dampen and smear together. Just as in the resonant gaps, where resonance does not dictate the optimal stiffness distribution, plates with a soft leading edge produce the smallest trailing edge amplitude, leading to the smallest power consumption. The benefits over a plate with uniformly distributed stiffness are modest when the plate is heaved, but pronounced when pitched, as we illustrate in figure~\ref{fig:powerchangeuni_opt}. In contrast, the benefits are great in the region dominated by Euler-Bernoulli behaviour, as being able to tune the stiffness distribution (and hence natural frequencies) allows us to avoid a resonant condition. 

\begin{figure}
  \begin{center}
  \includegraphics[width=\linewidth]{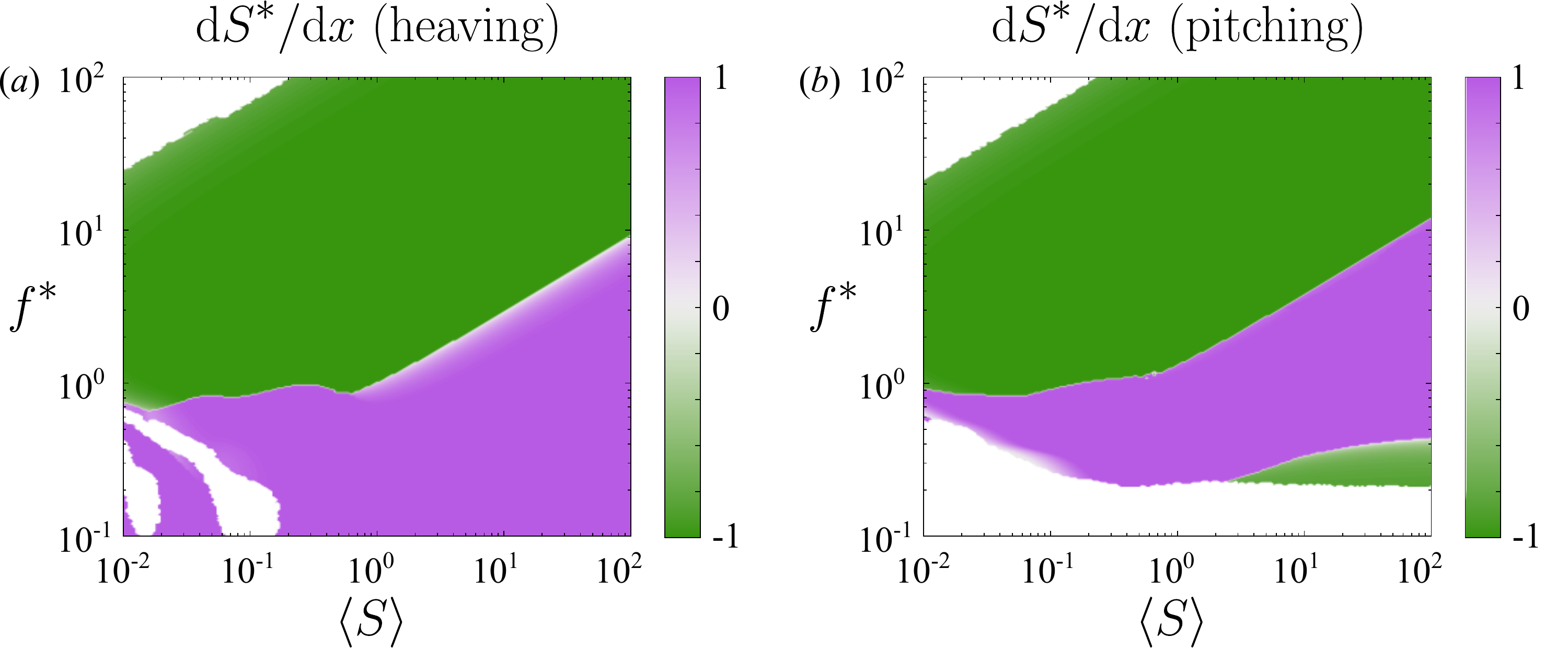}
  \end{center}
  \caption{Analog of figure~\ref{fig:thrust_opt_distr}, but for a linear
    stiffness distribution maximizing efficiency. }
  \label{fig:eta_opt_distr}
\end{figure}

\begin{figure}
  \begin{center}
  \includegraphics[width=\linewidth]{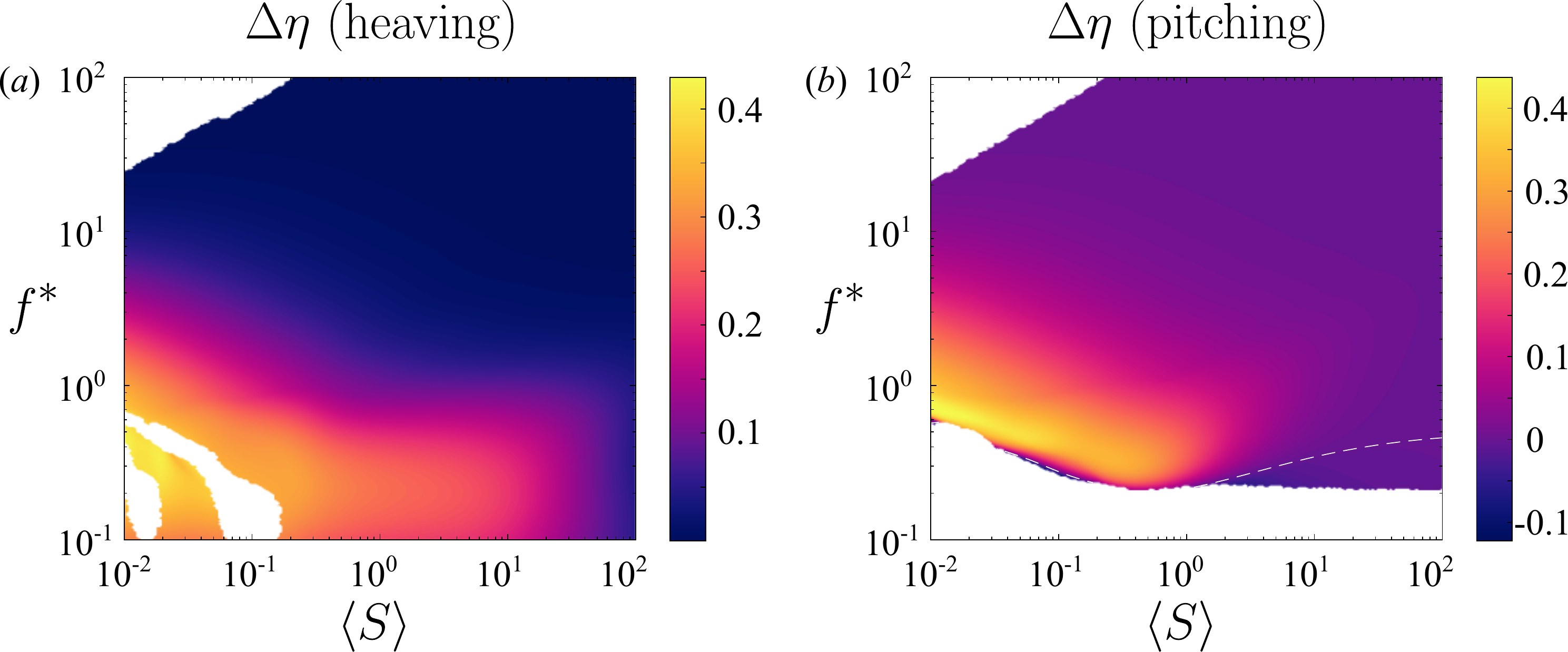}
  \end{center}
  \caption{Efficiency of a plate with the stiffness distribution shown in figure~\ref{fig:eta_opt_distr} relative to that of an equivalent rigid plate. Dashed white lines indicate where the flexible plate has the same efficiency as the equivalent rigid plate. }
  \label{fig:etadiff_opt}
\end{figure}

Considering the optimal stiffness distributions for maximizing thrust and minimizing power, it is not immediately clear what stiffness distribution will maximize efficiency. In the former case, resonance is sought after as a maximizer of thrust, and a stiff leading edge is preferred when resonance is not possible. In the latter case, resonance is avoided, and a soft leading edge is preferred when resonance is not possible. We present the optimal (efficiency-maximizing) stiffness distributions in figure~\ref{fig:eta_opt_distr}, with the attendant efficiency plotted in figure~\ref{fig:etadiff_opt}. When the plate is heaved, the optimizer sometimes converged to a solution with absolute efficiency greater than unity, with both thrust and power negative; we have whited out these cases. 

Unexpectedly, the frequency-stiffness plane is essentially divided into two zones: a lower zone where a plate with a soft leading edge is more efficient, and an upper zone where a plate with a stiff leading edge is more efficient. (For a pitching plate, a stiff leading edge is sometimes preferred near the zero-efficiency cutoff, where the thrust also crosses zero, since plates with a stiff leading edge produce more thrust than plates with a soft leading edge.) The boundary between the two zones changes qualitatively at $\langle S \rangle = 1$, i.e., when the behaviour changes from Euler-Bernoulli-dominated to flutter-dominated. In the Euler-Bernoulli region, the boundary is between the first and second natural frequencies, and runs parallel to them. In this region, elastic and added mass forces dominate the response, and the appropriate time scale is the bending time scale (see Section~\ref{sec:quiescent}). We therefore expect a boundary between regions to appear when the actuation and bending time scales are nearly equal, consistent with the results. In the flutter region, on the other hand, the boundary is near $f^* = 1$. In this region, lift and added mass forces dominate the response, and the appropriate time scale is the convective time scale. Although the boundary between zones in this region corresponds to the actuation and convective time scales being nearly equal, we caution that the limit $\langle S \rangle \rightarrow 0$ is a singular one. 

\begin{figure}
  \begin{center}
  \includegraphics[width=\linewidth]{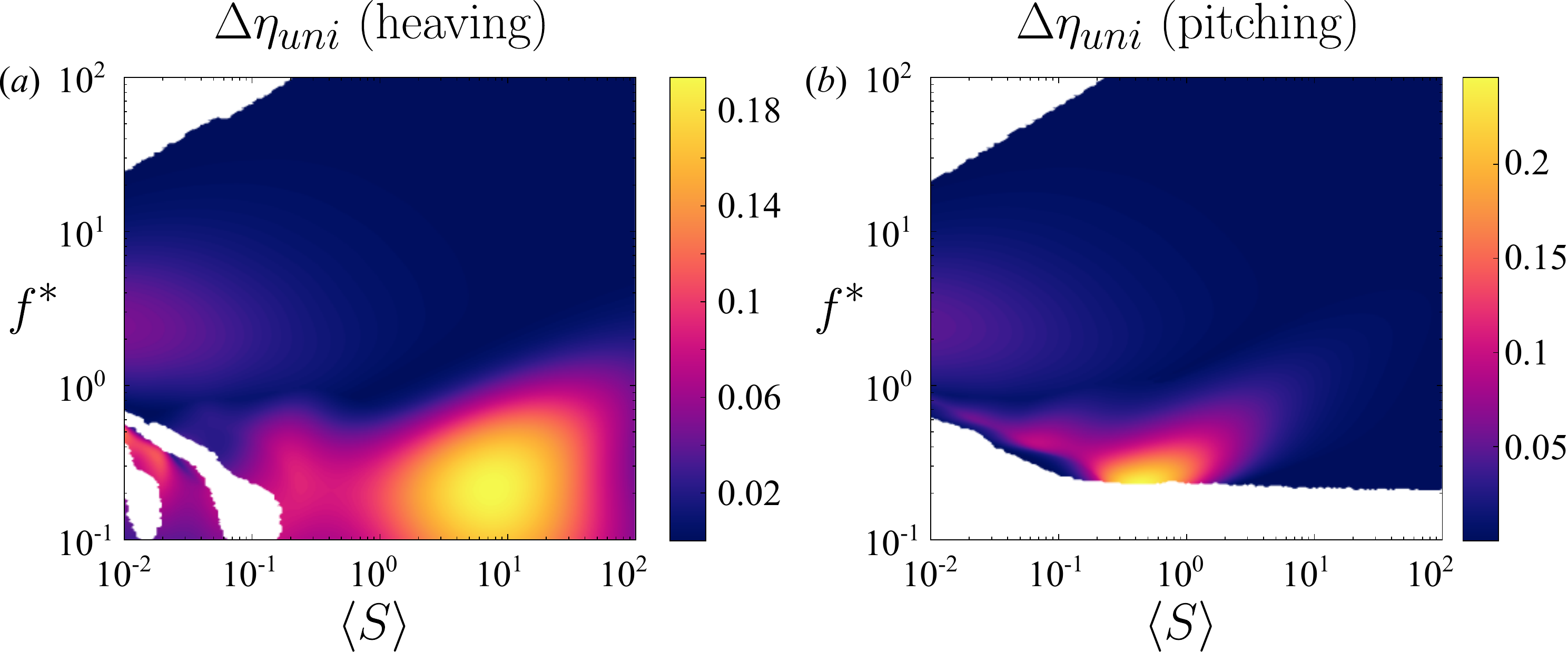}
  \end{center}
  \caption{Analog of figure~\ref{fig:etadiff_opt}, but compared to a plate with
    uniformly distributed stiffness instead of a rigid plate.}
  \label{fig:etadiffuni_opt}
\end{figure}

Why is a soft leading edge preferred in the lower zone, and a stiff leading edge preferred in the upper zone? To help answer this question, we appeal to the gains in efficiency made over a uniformly flexible plate, plotted in figure~\ref{fig:etadiffuni_opt}. The results show that meaningful efficiency gains are only made in the lower zone, i.e., by the plate with a flexible leading edge. We therefore focus on explaining this zone. 

\begin{figure}
  \begin{center}
  \includegraphics[width=\linewidth]{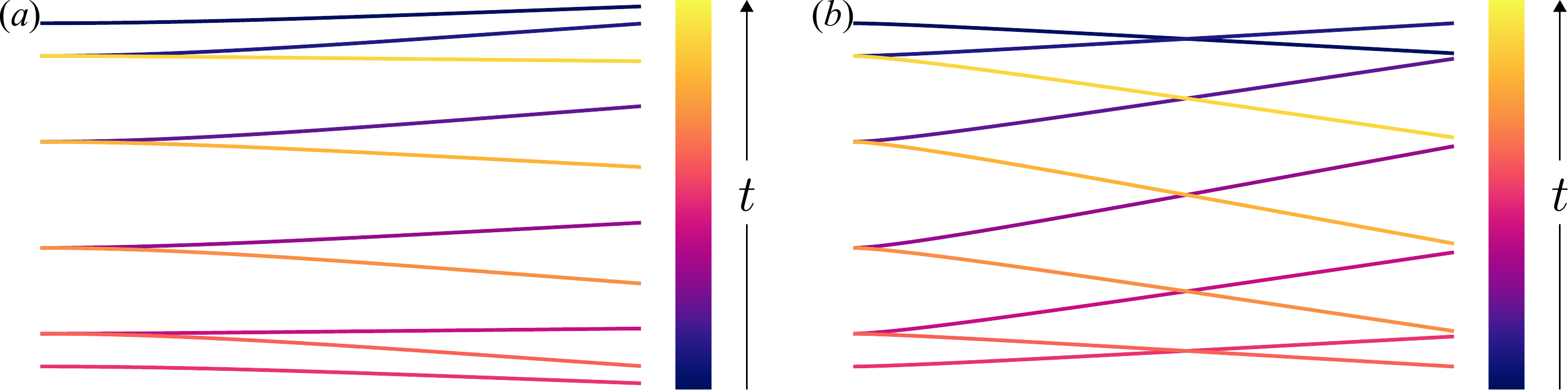}
  \end{center}
  \caption{Ten snapshots, evenly spaced in time, of a heaving plate with (a) uniform flexibility and (b) $\text{d}S/\text{d}x^* = 0.99$ (soft leading edge) comprising one period of motion. Both cases have the same heave amplitude, $\langle S \rangle = 8$, and $f^* = 0.2$. }
  \label{fig:kinheave}
\end{figure}

\begin{figure}
  \begin{center}
  \includegraphics[width=\linewidth]{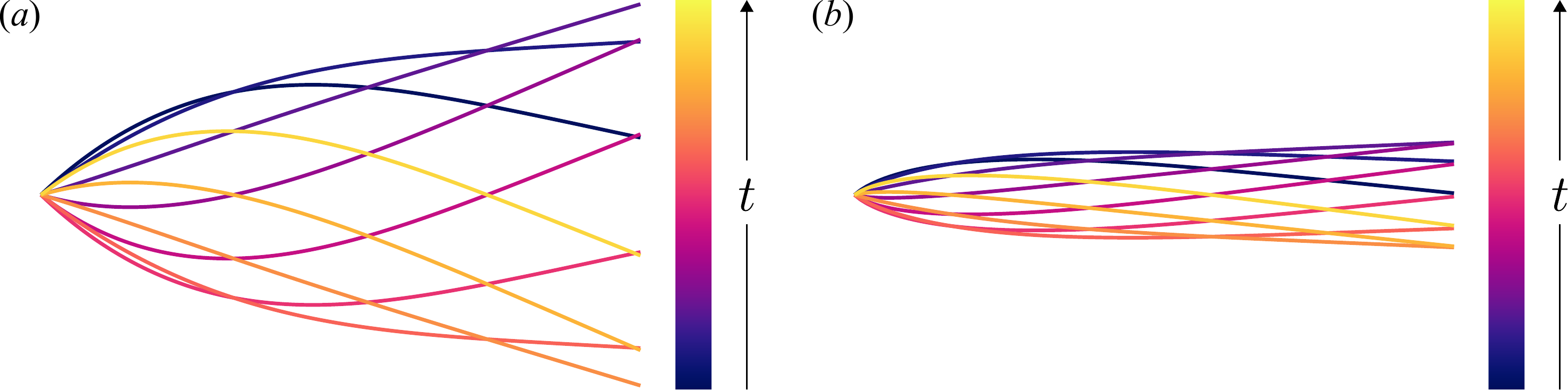}
  \end{center}
  \caption{Ten snapshots, evenly spaced in time, of a pitching plate with (a) uniform flexibility and (b) $\text{d}S/\text{d}x^* = 0.99$ (soft leading edge) comprising one period of motion. Both cases have the same pitch amplitude, $\langle S \rangle = 0.5$, and $f^* = 0.3$. }
  \label{fig:kinpitch}
\end{figure}

For uniformly flexible plates, meaningful gains in efficiency over rigid plates were made when flutter modes appeared. The flutter modes induce travelling wave kinematics in the actuated plate, which are known to be efficient \citep{wu1961swimming}. When the leading edge is soft, we saw that the natural frequencies decrease compared to a uniformly flexible plate. Moreover, flutter modes appear at higher mean stiffness ratios for plates with a soft leading edge than for uniformly flexible plates. Consequently, travelling wave kinematics can be induced in plates with a soft leading edge at higher values of the mean stiffness ratio than for uniformly flexible plates. Indeed, the area in the frequency-stiffness plane where the most significant gains in efficiency are made in going from a uniform stiffness distribution to an optimal stiffness distribution is where flutter modes are present for a plate with a soft leading edge but not present for a uniformly flexible plate (at least when the plate is heaved). When the plate is pitched, the region of greatest efficiency gains is shifted since a pitching plate produces net drag at low enough frequency. The physical reason for the efficiency gains, however, is unchanged: the plate with a soft leading edge has nice travelling wave kinematics in that region. These kinematics are shown in figure~\ref{fig:kinheave} for a heaving plate and figure~\ref{fig:kinpitch} for a pitching plate and contrasted to the kinematics of a uniformly flexible plate. 

\subsection{Quadratic stiffness distributions}
\label{sec:quad}

\begin{figure}
  \begin{center}
  \includegraphics[width=0.4\linewidth]{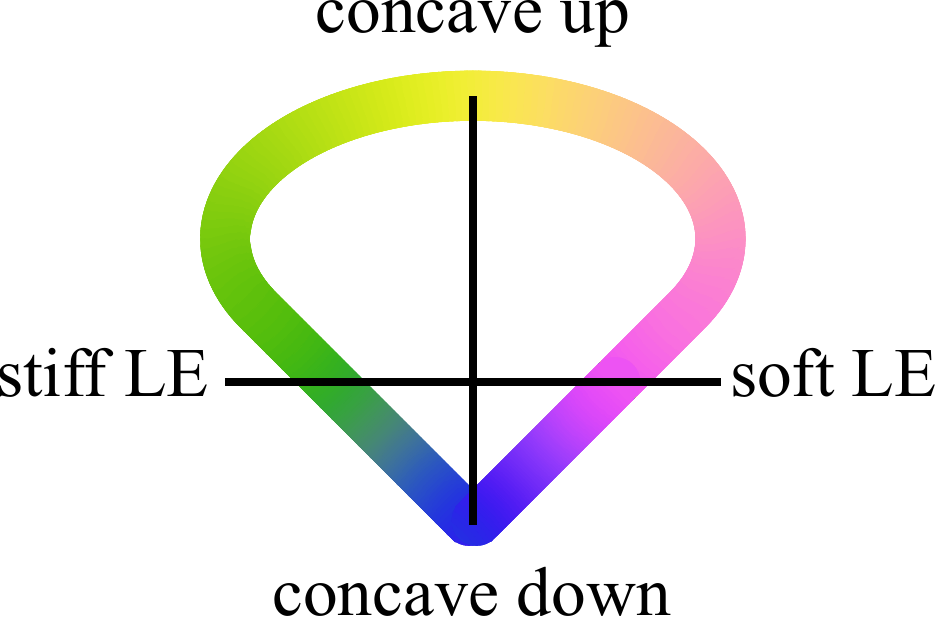}
  \end{center}
  \caption{Colour coding of the boundary of the feasible set in figure~\ref{fig:feas}.}
  \label{fig:legend}
\end{figure}

We now add an additional degree of freedom, allowing the stiffness distribution
to vary quadratically along the chord. As it turns out, all of the optimal
distributions for the parameter values studied here lie near the boundary of the
feasible set depicted in figure~\ref{fig:feas}. (When performing the
optimization, we slightly shrunk the feasible set in order to avoid problems
associated with the stiffness being zero somewhere along the chord.) We may
therefore represent the optimal quadratic stiffness distributions by a single
parameter describing the location along the boundary, as shown in
figure~\ref{fig:legend}. As before, stiffness distributions with a stiff leading
edge are coded as green and those with a soft leading edge are coded as
purple. In addition, stiffness distributions that are concave up (edges stiffer
than the interior) are coded as yellow and those that are concave down (edges
softer than the interior) are coded as blue. Distributions on the horizontal
line are linear, and distributions on the vertical line are symmetric about the
mid-chord.

\begin{figure}
  \begin{center}
  \includegraphics[width=\linewidth]{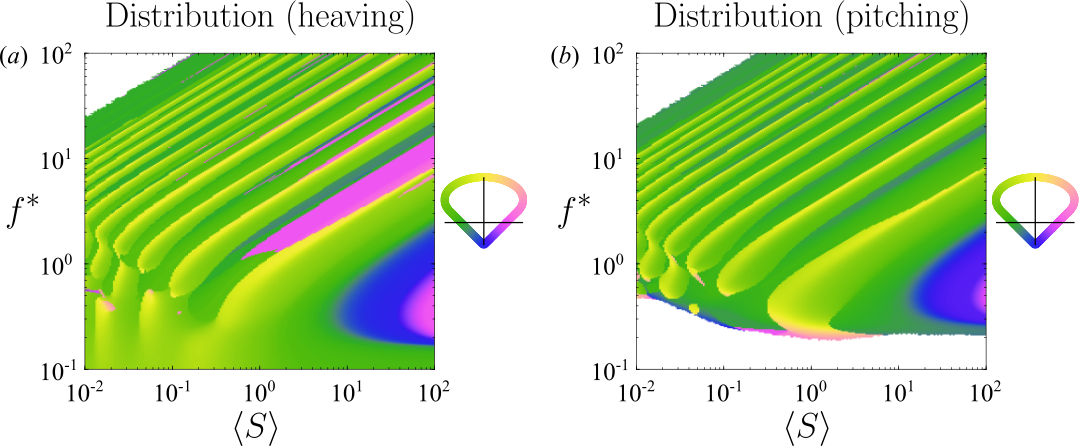}
  \end{center}
  \caption{Thrust-maximizing quadratic stiffness distribution as a function of reduced frequency $f^*$ and mean stiffness ratio $\langle S \rangle$ for a (a) heaving and (b) pitching plate with $R \equiv 0.01$. Under-resolved areas and areas that produce negative thrust have been whited out. }
  \label{fig:thrust_optquad_distr}
\end{figure}

\begin{figure}
  \begin{center}
  \includegraphics[width=\linewidth]{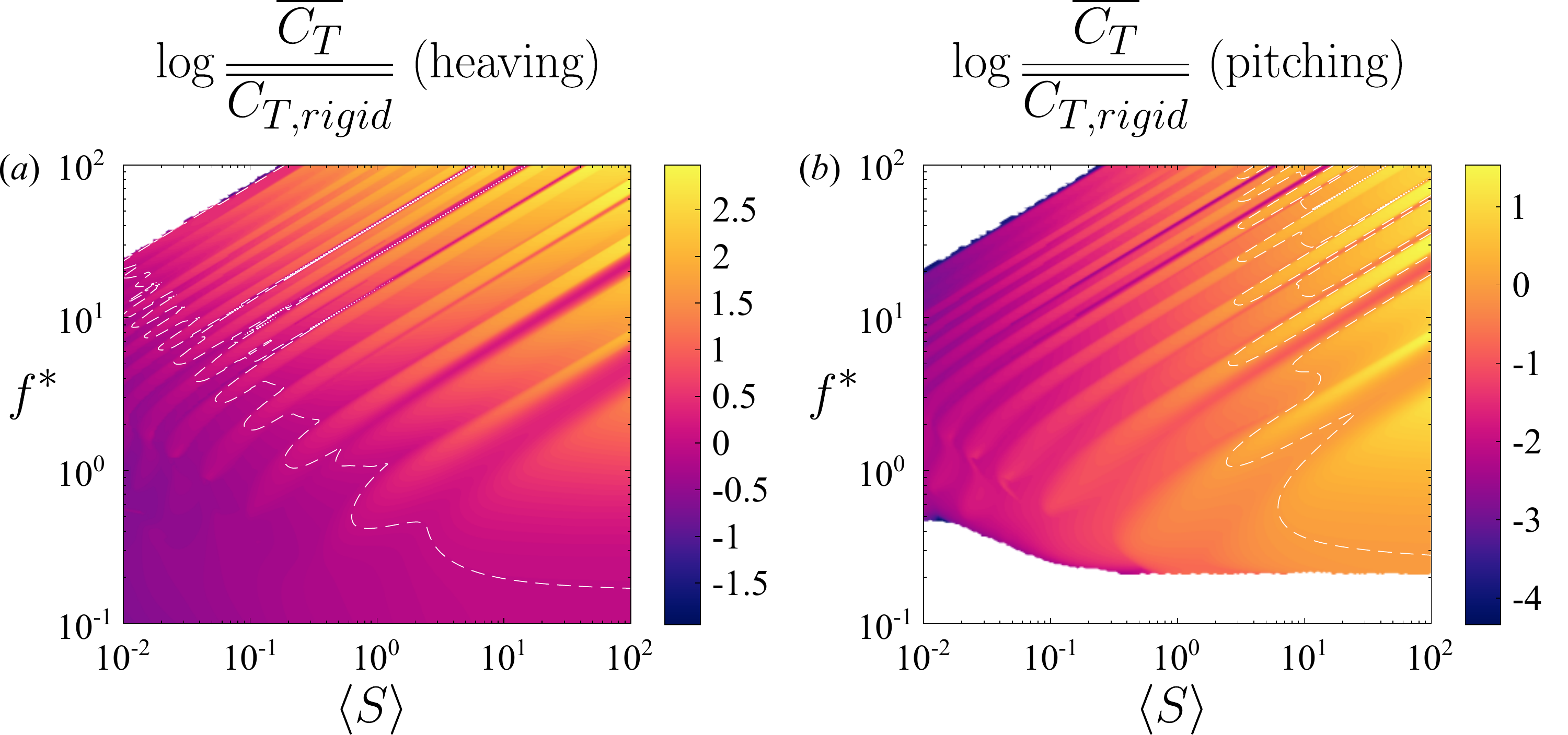}
  \end{center}
  \caption{Thrust coefficient of a plate with the stiffness distribution shown in figure~\ref{fig:thrust_optquad_distr} relative to that of an equivalent rigid plate. Dashed white lines indicate where the flexible plate has the same thrust coefficient as the equivalent rigid plate.}
  \label{fig:thrust_optquad}
\end{figure}

In figure~\ref{fig:thrust_optquad_distr}, we have plotted the optimal
(thrust-maximizing) quadratic stiffness distribution, with the attendant optimal
mean thrust coefficient plotted in figure~\ref{fig:thrust_optquad}. The overall
trends are comparable to those for linear stiffness distributions. In the region
dominated by Euler-Bernoulli modes, the optimal stiffness distribution at a
given reduced frequency and mean stiffness ratio is the one that has a natural
frequency at that frequency of actuation. The additional degree of freedom in
the stiffness distribution gives more freedom to tune the natural frequency of
the plate, thereby broadening the resonant response and narrowing the resonant
gaps. As for linear distributions, a stiff leading edge is preferred when a
resonant condition cannot be reached, both in the region dominated by
Euler-Bernoulli modes and the region dominated by flutter modes. When
distributions with a stiff or soft leading edge have the same natural frequency,
a stiff leading edge is again preferred. With the additional degree of freedom
in stiffness distribution, the natural frequencies of distributions with stiff
leading edges are able to cover a larger portion of the frequency-stiffness
plane than linear distributions. As a result, a larger portion of the
frequency-stiffness plane is green in figure~\ref{fig:thrust_optquad_distr}
(cf.\ figure~\ref{fig:thrust_opt_distr}); this is especially evident for pitching
motions, where distributions with a soft leading edge have almost entirely
disappeared.

We note that much of the frequency-stiffness plane has a yellow tint, reflecting that a positive quadratic component of the stiffness distribution enhances the thrust. The reason is quite simple: a positive quadratic component sacrifices the stiffness of the interior of the plate to increase the stiffness of the edges. In other words, the leading edge can be made even stiffer than with just a linear stiffness distribution while maintaining the same mean stiffness. The quadratic component of the stiffness distribution allows us to concentrate the stiffness toward the leading edge, which we have seen enhances thrust. The trends for quadratic stiffness distributions are the same as for linear stiffness distributions. We posit that even higher-order distributions would tend to further concentrate the stiffness at the leading edge. 

\begin{figure}
  \begin{center}
  \includegraphics[width=\linewidth]{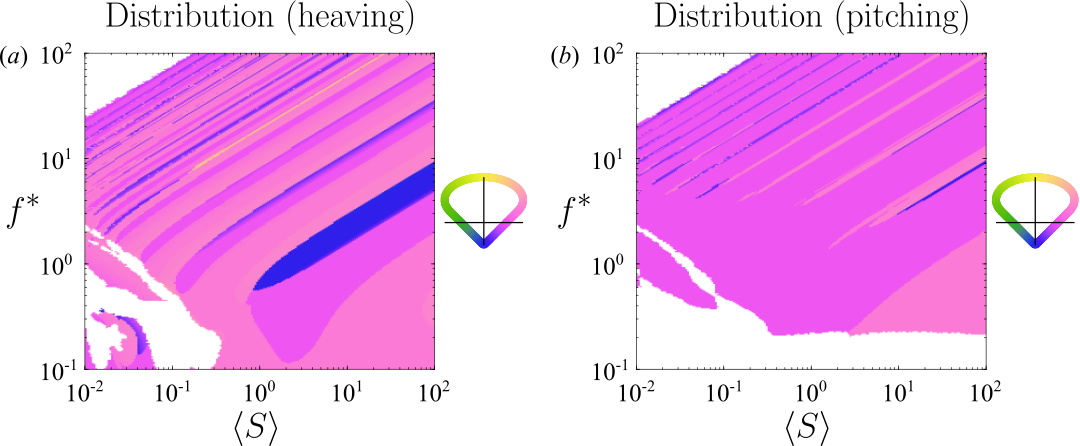}
  \end{center}
  \caption{Analog of figure~\ref{fig:thrust_optquad_distr}, but for a quadratic
    stiffness distribution minimizing power.}
  \label{fig:power_optquad_distr}
\end{figure}

\begin{figure}
  \begin{center}
  \includegraphics[width=\linewidth]{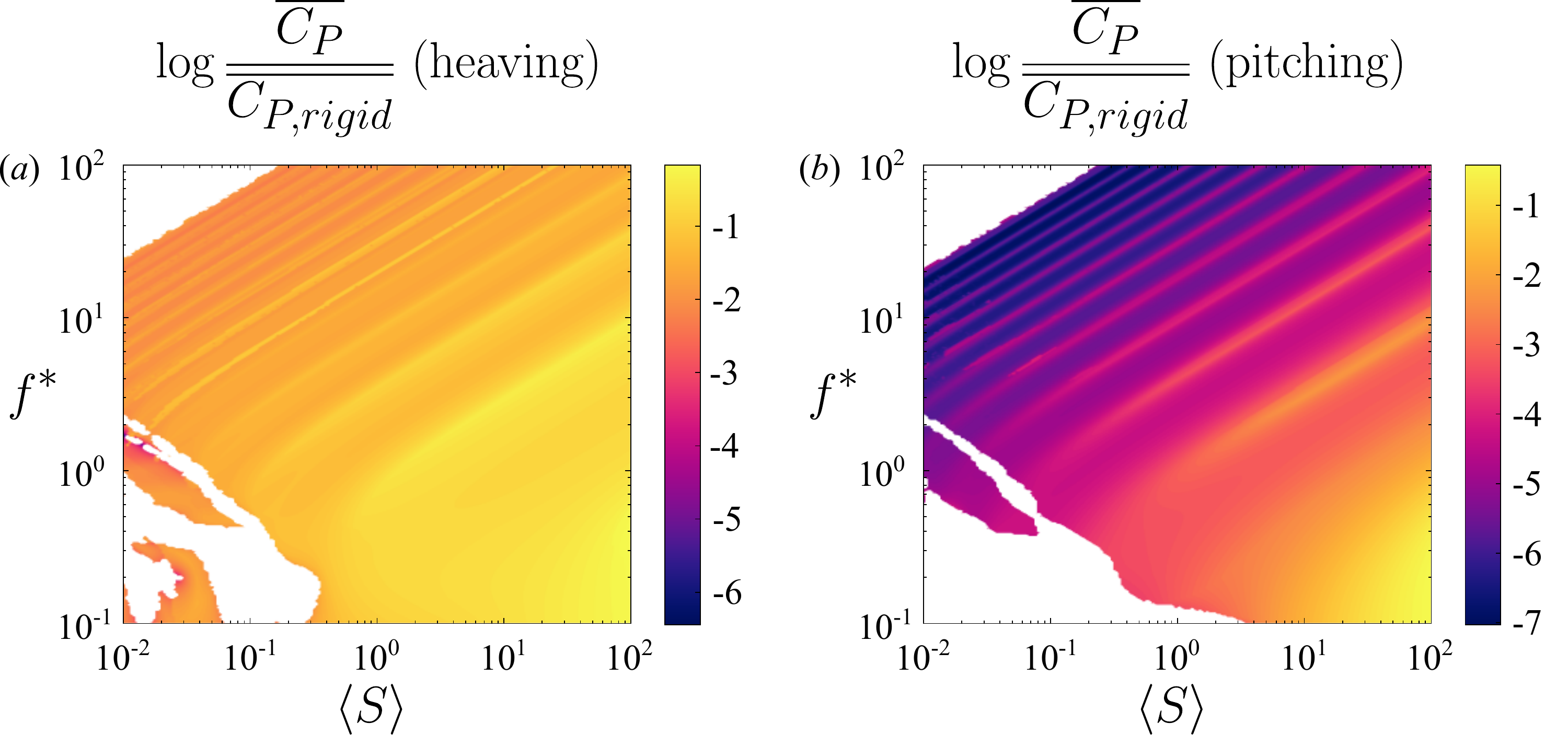}
  \end{center}
  \caption{Power coefficient of a plate with the stiffness distribution shown in figure~\ref{fig:power_optquad_distr} relative to that of an equivalent rigid plate. Dashed white lines indicate where the flexible plate has the same power coefficient as the equivalent rigid plate.}
  \label{fig:power_optquad}
\end{figure}

When minimizing the power, the results mirror those for linear stiffness distributions. The optimal quadratic stiffness distribution is plotted in figure~\ref{fig:power_optquad_distr}, with the attendant optimal mean power coefficient plotted in figure~\ref{fig:power_optquad}. Essentially, a soft leading edge is preferred unless it creates a condition of resonance. In other words, the results are opposite of those when maximizing thrust. We note that much of the frequency-stiffness plane has a pink tint, reflecting a positive quadratic component of the stiffness distribution. The positive quadratic component allows us to concentrate the stiffness toward the trailing edge, making the leading edge softer than a linear distribution could, thereby further decreasing power consumption. There are also portions of the frequency-stiffness plane that are blue, with the stiffness being distributed symmetrically about the mid-chord and concentrated toward the interior of the plate. The blue regions largely replace regions where a uniform stiffness distribution was preferred for linear distributions, and appear in the region dominated by Euler-Bernoulli modes. In the blue regions, stiffness distributions with the stiffness concentrated away from the leading edge apparently have a natural frequency present in those regions, which would increase power consumption. The symmetric, concave-down distribution is the best option as it still softens the leading edge compared to the uniform distribution (although the leading and trailing edges are equally stiff).  

Overall, the trends are the same as for linear stiffness distributions: concentrate stiffness away from the leading edge while avoiding resonance. For higher-order stiffness distributions, we posit that the preferred distribution will continue to be the one that most effectively concentrates stiffness away from the leading edge. Once the order is high enough, it may be that at each point in the frequency-stiffness plane there exists a distribution with a soft leading edge without a natural frequency at that point in the frequency-stiffness plane. 

\begin{figure}
  \begin{center}
  \includegraphics[width=\linewidth]{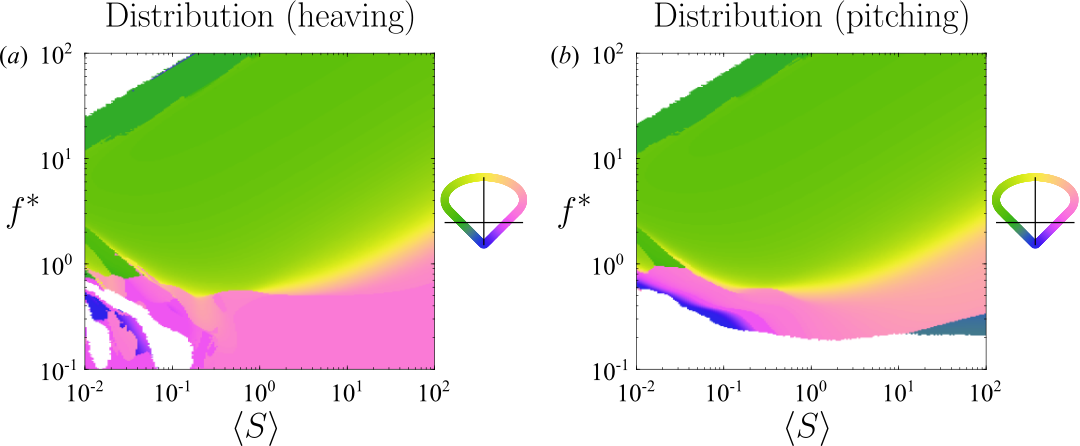}
  \end{center}
  \caption{Analog of figure~\ref{fig:thrust_optquad_distr}, but for a quadratic stiffness distribution maximizing efficiency. }
  \label{fig:eta_optquad_distr}
\end{figure}

\begin{figure}
  \begin{center}
  \includegraphics[width=\linewidth]{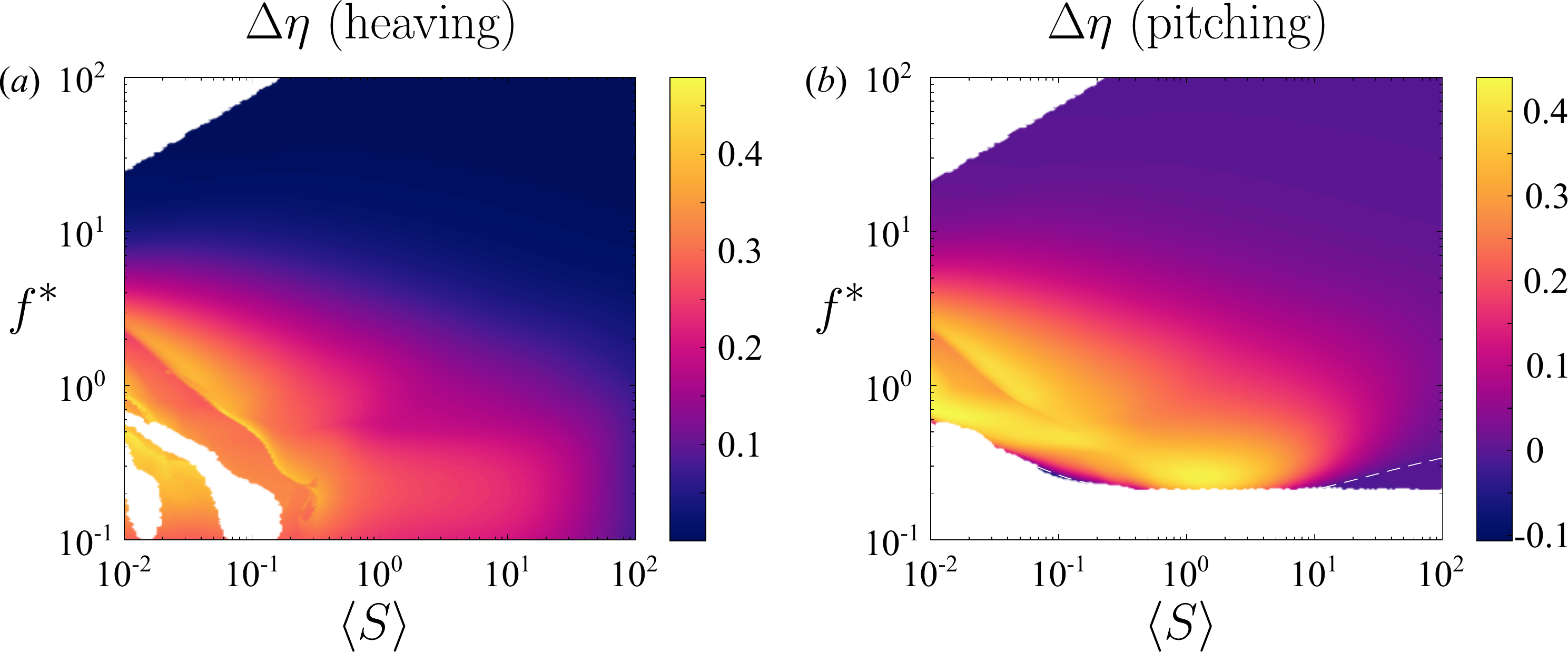}
  \end{center}
  \caption{Efficiency of a plate with the stiffness distribution shown in figure~\ref{fig:eta_optquad_distr} relative to that of an equivalent rigid plate. Dashed white lines indicate where the flexible plate has the same efficiency as the equivalent rigid plate.}
  \label{fig:etadiff_optquad}
\end{figure}

Since a stiff leading edge generally maximizes thrust production and a soft leading edge generally minimizes power consumption, it is not immediately clear what stiffness distribution will maximize efficiency. The efficiency-maximizing stiffness distributions are plotted in figure~\ref{fig:eta_optquad_distr}, with the attendant efficiency plotted in figure~\ref{fig:etadiff_optquad}. When the plate is heaved, the optimizer sometimes converged to a solution with absolute efficiency greater than unity, with both thrust and power negative; we have whited out these cases. 

The results mirror those for linear stiffness distributions, with the frequency-stiffness plane essentially divided into two zones: a lower zone where a plate with a soft leading edge is more efficient, and an upper zone where a plate with a stiff leading edge is more efficient. For a pitching plate, there is a small region near the zero-thrust cutoff where a plate with a stiff leading edge is more efficient. For $\langle S \rangle \gtrsim 1$, boundary between the two zones runs parallel to the natural frequencies and is quite broad compared to the same boundary for linear stiffness distributions. For $\langle S \rangle \lesssim 1$, the boundary between the two zones is irregular but still quite sharp. In most of the frequency-stiffness plane, the optimal stiffness distributions have a positive quadratic component, meaning that concentrating stiffness towards the edges is beneficial for efficiency. This effect was also seen when maximizing thrust and minimizing power, so we posit that it will continue to hold for higher-order stiffness distributions. 

\begin{figure}
  \begin{center}
  \includegraphics[width=\linewidth]{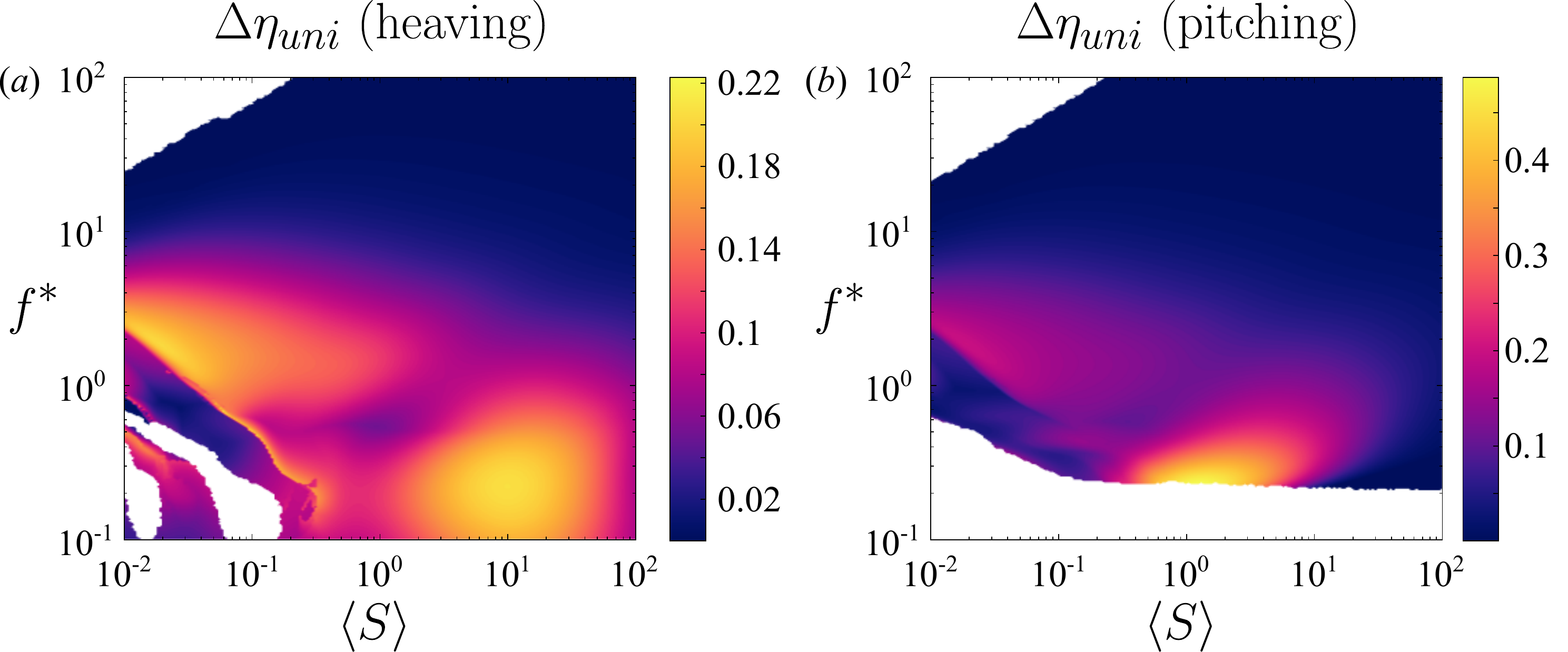}
  \end{center}
  \caption{Analog of figure~\ref{fig:etadiff_optquad}, but compared to a plate with uniformly distributed stiffness instead of a rigid plate.}
  \label{fig:etadiffuni_optquad}
\end{figure}

Again, only certain portions of the frequency-stiffness plane enjoy meaningful gains in efficiency over a uniformly flexible plate, shown in figure~\ref{fig:etadiffuni_optquad}. With the addition of the quadratic component to the stiffness distribution, a significant portion of the frequency-stiffness plane where a stiff leading edge is preferred enjoys meaningful gains in efficiency over a uniformly flexible plate, whereas this was not the case for linear stiffness distributions. Although we do not show it here for brevity, the regions with meaningful gains in efficiency are those where efficient travelling wave kinematics are induced by flutter modes. 

\subsection{Finite Reynolds number effects}
\label{sec:re}

We take the opportunity to briefly speculate on the effects of streamwise drag. Having an offset drag in the system will move where the net thrust transitions from being negative to positive to greater frequencies. Distributions with stiffness concentrated toward the leading edge will still produce the greatest thrust, and the results for power consumption will also be unaffected. 

Efficiency is a different story, though. Drag can create peaks in efficiency nearby where the net thrust transitions from being negative to positive \citep{floryan2017scaling, floryan2018efficient}. The presence of drag also makes the efficiency quite sensitive to changes in the system. These two effects are present for both rigid and flexible plates. 

For flexible plates, an additional effect emerges. As shown in figure~\ref{fig:etadiff_uni}, the efficiency of flexible plates does not have any resonant peaks in the inviscid small-amplitude regime, with the resonant peaks in thrust and power cancelling each other exactly. Upon adding drag to the system, however, resonant peaks in efficiency emerge \citep{floryan2018clarifying}. The explanation is straightforward. In moving from a non-resonant to a resonant condition in a system without drag, the mean thrust and power coefficients effectively scale up by some factor $a > 1$. Since the efficiency is the ratio of the two, the factor $a$ appears in the numerator and denominator, cancels, and there is no resonant peak. When drag is present, it reduces the baseline non-resonant efficiency compared to the system without drag. In moving to a resonant condition, the thrust and power scale up by the factor $a$, but the drag does not. The net thrust, therefore, scales up by a factor greater than $a$, so the efficiency increases at resonance. This effect creates local maxima in efficiency at resonance. Since drag affects the system at first order, this effect of streamwise drag inducing resonant peaks in efficiency should be robust to nonlinearities present at finite amplitudes. 

The presence of streamwise drag will, therefore, have two effects on the efficiency-maximizing stiffness distributions. The first is that it will wipe out the stiffness distributions with a soft leading edge at low frequencies. At low frequencies, the distributions with a soft leading edge that are highly efficient also produce very little thrust. The drag will be comparable in magnitude to the thrust produced by plates with a soft leading edge, causing the net thrust (and therefore the efficiency) to plummet. Since plates with a stiff leading edge produce greater thrust, they are more robust to the effects of drag and will be favoured in the presence of drag over plates with a soft leading edge. 

The second effect occurs far from where the net thrust transitions between negative and positive. Since drag induces a resonant behaviour in efficiency, the efficiency-maximizing stiffness distribution will take advantage of this resonant effect. Consequently, the efficiency-maximizing stiffness distribution will tend towards the thrust-maximizing stiffness distribution. Altogether, the presence of drag will make the efficiency-maximizing stiffness distribution tend towards the thrust-maximizing stiffness distribution everywhere in the frequency-stiffness plane. 

\section{Conclusions}
\label{sec:conc}

In this work, we studied a linear inviscid model of a passively flexible swimmer with distributed flexibility, valid for small-amplitude, low-frequency motions where there is no separation. We were careful to separate the effects due to mean stiffness from those due to the distribution of stiffness. The frequencies of actuation and mean stiffness ratios we considered spanned a large range, while the mass ratio was mostly fixed to a low value representative of swimmers. For low values of the mean mass ratio, the spatial distribution of mass matters little, but for mass ratios of order unity and higher the spatial distribution may matter. The results presented in this work are therefore applicable to swimmers, and care should be taken in extending the results to fliers. 

Qualitatively, the trailing edge deflection, thrust coefficient, power coefficient, and efficiency vary similarly with mean stiffness and frequency for a plate with distributed flexibility as they do for a plate with uniform flexibility. The trailing edge deflection, thrust coefficient, and power coefficient showed sharp ridges of resonant behaviour for reduced frequencies $f^* > 1$ and stiffness ratios $S > 1$, where Euler-Bernoulli modes govern the dynamics. For $f^* < 1$ and $S < 1$, however, the resonant peaks smeared together. The efficiency, on the other hand, did not show resonant peaks anywhere in the frequency-stiffness plane, instead showing a broad region of high values for $f^* < 1$ and $S < 1$, where flutter modes govern the dynamics and induce efficient travelling wave kinematics. 

Important quantitative differences between plates with distributed and uniform flexibility exist, however, which we elucidated by optimizing the stiffness distribution. To maximize thrust, the stiffness distribution should be tuned so that a natural frequency coincides with the frequency of actuation, triggering a resonant response; if this is not possible, then stiffness should be concentrated towards the leading edge. To minimize power, the opposite conclusions hold: avoid resonance, or concentrate stiffness away from the leading edge if resonance cannot be avoided. To maximize efficiency, the stiffness should be concentrated towards the leading edge for high frequencies and away from the leading edge for low frequencies. Meaningful gains in efficiency over a uniformly flexible plate were only made for low frequencies. Here, concentrating stiffness away from the leading edge induced efficient travelling wave kinematics

Lastly, we speculated on the effects of a finite Reynolds number in the form of streamwise drag. Streamwise drag adds an offset drag to the system, which shifts the zero-thrust cutoff to higher frequencies and creates resonant peaks in the efficiency that are not present in the inviscid system. Consequently, efficiency-maximizing distributions of flexibility will tend towards thrust-maximizing distributions everywhere in the frequency-stiffness plane for real systems with drag; i.e., they will concentrate stiffness towards the leading edge unless a resonant response can be triggered by concentrating stiffness away from the leading edge. We note that animals tend to concentrate stiffness towards the leading edge. \\

This work was supported by ONR Grant N00014-14-1-0533 (Program Manager R. Brizzolara).

Declaration of interests. The authors report no conflict of interest.

\appendix

\section{Method of solution}
\label{sec:sol}
We may write the deflection as a Chebyshev series with time-varying coefficients,
\begin{equation}
  \label{eq:sol1}
  Y(x,t) = \displaystyle\frac{1}{2}\beta_0(t) + \sum_{k=1}^\infty \beta_k(t) T_k(x),
\end{equation}
where $T_k(x) = \cos(k \arccos x)$ is the Chebyshev polynomial of degree $k$. For coefficients that vary sinusoidally in time, the solution to the flow is given in \citet{wu1961swimming}; we repeat the basics of that analysis in the proceeding text. 

Represent two-dimensional physical space $(x,y)$ by the complex plane $z = x + \mathrm{i}y$, where $\mathrm{i} = \sqrt{-1}$. There exists a complex potential $F(z,t) = \phi(z,t) + \mathrm{i}\psi(z,t)$, with $\phi$ and $\psi$ harmonic conjugates, that is analytic in $z$ and related to the complex velocity $w = u - \mathrm{i}v$ through the momentum equation by
\begin{equation}
  \label{eq:sol2}
  \frac{\partial F}{\partial z} = \frac{\partial w}{\partial t} + \frac{\partial w}{\partial z}.
\end{equation}
We use the conformal transformation
\begin{equation}
  \label{eq:sol3}
  z = \frac{1}{2}\left( \zeta + \frac{1}{\zeta}\right)
\end{equation}
to map physical space in the $z$-plane to the exterior of the unit circle in the $\zeta$-plane. This transformation maps the plate onto the unit circle. The complex potential can be represented by a multipole expansion
\begin{equation}
  \label{eq:sol4}
  F(\zeta,t) = \phi(\zeta,t) + \mathrm{i}\psi(\zeta,t) = \mathrm{i} \left( \frac{a_0(t)}{\zeta + 1} + \sum_{k=1}^\infty \frac{a_k(t)}{\zeta^k}\right).
\end{equation}
Evaluating on the unit circle $\zeta = \mathrm{e}^{\mathrm{i}\theta}$ gives
\begin{equation}
  \label{eq:sol5}
  \left. \begin{array}{ll}
  \phi(\zeta = \mathrm{e}^{\mathrm{i}\theta},t) = \displaystyle\frac{1}{2}a_0(t) \tan\frac{\theta}{2} + \sum_{k=1}^\infty a_k(t) \sin k\theta, \\[8pt]
  \psi(\zeta = \mathrm{e}^{\mathrm{i}\theta},t) = \displaystyle\frac{1}{2}a_0(t) + \sum_{k=1}^\infty a_k(t) \cos k\theta.
  \end{array}\right\}
\end{equation}
In physical space, on the surface of the plate we have
\begin{equation}
  \label{eq:sol6}
  \left. \begin{array}{ll}
  \phi(z = x,t) = \Phi^\pm(x,t)= \pm\displaystyle\frac{1}{2}a_0(t)\sqrt{\frac{1-x}{1+x}} \pm \sum_{k=1}^\infty a_k(t) \sin k\theta, \\[8pt]
  \psi(z = x,t) = \Psi(x,t)= \displaystyle\frac{1}{2}a_0(t) + \sum_{k=1}^\infty a_k(t) T_k(x),
  \end{array}\right\}
\end{equation}
where we have used $x = \cos\theta$; $\psi$ has equal values on the top and bottom since it is even in $\theta$, whereas $\phi$ is odd in $\theta$ and thus has a discontinuity in physical space. 

At this point, it is convenient to explicitly write out the sinusoidal-in-time dependence of the coefficients,
\begin{equation}
  \label{eq:sol6b}
  \left. \begin{array}{ll}
  Y(x,t) = \Real\{\mathrm{e}^{\mathrm{i} \sigma t} \hat Y(x)\}, \\[8pt]
  \beta_k(t) = \Real\{\mathrm{e}^{\mathrm{i} \sigma t} \hat \beta_k\}, \\[8pt]
  a_k(t) = \Real\{\mathrm{e}^{\mathrm{i} \sigma t} \hat a_k\}.
  \end{array}\right\}
\end{equation}
The no-penetration condition can be written as
\begin{equation}
  \label{eq:sol7}
  \frac{\partial \psi}{\partial x}|_{y=0} = -\left(\frac{\partial}{\partial t} + \frac{\partial}{\partial x}\right)^2 Y,
\end{equation}
which simplifies to 
\begin{equation}
  \label{eq:sol8}
  D\Psi = \Real\{-(\mathrm{i}\sigma + D)^2 \hat Y\},
\end{equation}
where $D = \mathrm{d}/\mathrm{d}x$. Given $\hat Y$, this equation allows us to solve for all $\hat a_k$ except $\hat a_0$. To solve for $\hat a_0$, we begin by writing the vertical velocity on the surface of the plate as
\begin{equation}
  \label{eq:sol9}
  v(z = x,t) = \Real\{\mathrm{e}^{\mathrm{i}\sigma t}\hat V(x)\} = \Real\left\{\mathrm{e}^{\mathrm{i}\sigma t}\left(\frac{1}{2}\hat V_0 + \sum_{k=1}^\infty \hat V_k T_k(x)\right)\right\}.
\end{equation}
The no-penetration condition can then be written as
\begin{equation}
  \label{eq:sol10}
  \hat V = (\mathrm{i}\sigma + D)\hat Y.
\end{equation}
The coefficient $\hat a_0$ is given by
\begin{equation}
  \label{eq:sol11}
  \hat a_0 = -C(\mathrm{i}\sigma)(\hat V_0 + \hat V_1) + \hat V_1,
\end{equation}
where
\begin{equation}
  \label{eq:sol12}
  C(\mathrm{i}\sigma) = \frac{K_1(\mathrm{i}\sigma)}{K_0(\mathrm{i}\sigma) + K_1(\mathrm{i}\sigma)}
\end{equation}
is the Theodorsen function, and $K_\nu$ is the modified Bessel function of the second kind of order $\nu$. The expression for $\hat a_0$ is derived in \citet{wu1961swimming}.

With all of the $\hat a_k$ known, the pressure difference across the plate can be written as
\begin{equation}
  \label{eq:sol13}
  \Delta p(x,t) = \Real\{\mathrm{e}^{\mathrm{i}\sigma t}\hat P(x)\} = \Real\left\{\mathrm{e}^{\mathrm{i}\sigma t}\left(\hat a_0\sqrt{\frac{1-x}{1+x}} + 2\sum_{k=1}^\infty \hat a_k \sin k\theta\right)\right\}.
\end{equation}
We note that the pressure difference depends linearly on the deflection $\hat Y$. 

Altogether, given the deflection $\hat Y$, we may calculate the coefficients $\hat a_k$. The coefficients $\hat a_k$ are used to calculate the pressure difference across the plate, which alters the deflection of the plate via~\eqref{eq:prob3}. The coupled fluid-structure problem must be solved numerically.

\subsection{Numerical method}
\label{sec:num}
Substituting the Chebyshev series~\eqref{eq:sol1} into the Euler-Bernoulli equation~\eqref{eq:prob3} gives a fourth-order differential equation for $\hat Y$,
\begin{equation}
  \label{eq:num1}
  -2\sigma^2 R\hat Y + \frac{2}{3}D^2(S D^2 \hat Y) = \hat P.
\end{equation}
The corresponding boundary conditions~\eqref{eq:prob6} are re-written as
\begin{equation}
  \label{eq:num2}
  \hat Y(-1) = h_0,\quad \hat Y_x(-1) = \theta_0,\quad \hat Y_{xx}(1) = 0,\quad \hat Y_{xxx}(1) = 0,
\end{equation}
where $h_0$ and $\theta_0$ are the heaving and pitching amplitudes at the leading edge, respectively. We re-iterate that the pressure difference across the plate $\hat P$ is a linear function of the deflection $\hat Y$, and so~\eqref{eq:num1}--\eqref{eq:num2} give a linear, homogeneous boundary value problem for $\hat Y$. When solving for the deflection $\hat Y$, all infinite series are truncated to the upper limit $N$. 

The numerical method to solve the boundary value problem is given in \citet{moore2017fast}. The method is a pseudo-spectral Chebyshev scheme that uses Gauss-Chebyshev points. The method is fast (\textit{O}$(N \log N)$) and accurate, avoiding errors typically encountered when using Chebyshev methods to solve high-order differential equations by pre-conditioning the system with continuous operators. Quadrature formulas for the thrust and power coefficients in~\eqref{eq:prob7} and~\eqref{eq:prob8} are also given in \citet{moore2017fast}.

\section{Eigenvalues of the system}
\label{sec:eig}
Here, we seek to determine the natural response of a flexible plate whose leading edge is held clamped in an oncoming flow \citep{alben2008flapping, michelin2009resonance, eloy2007flutter}. This amounts to finding the eigenvalues and eigenvectors of the system~\eqref{eq:prob3} with homogeneous boundary conditions ($h(t) \equiv 0$ and $\theta(t) \equiv 0$). To do so, quantities that were previously written as Fourier-Chebyshev expansions (the deflection, complex potential, and velocity) are now written as Chebyshev series with time-varying coefficients. Following the preceding analysis, we arrive at the following equations:
\begin{align}
  \label{eq:eig1}
  2RY_{tt} + \frac{2}{3}(SY_{xx})_{xx} = \Delta p, \\
  Y(x,t) = \frac{1}{2}\beta_0(t) + \sum_{k=1}^\infty \beta_k(t)T_k(x), \\
  \Delta p(x,t) = a_0(t)\sqrt{\frac{1-x}{1+x}} + 2\sum_{k=1}^\infty a_k(t) \sin k\theta, \\
  \sum_{k=1}^\infty a_kT_k' = -\frac{1}{2}\ddot{\beta_0} - \sum_{k=1}^\infty \left[\ddot{\beta_k}T_k + 2 \dot{\beta_k}T_k' + \beta_kT_k''\right],
\end{align}
where a dot denotes differentiation with respect to $t$ and a prime denotes differentiation with respect to $x$. 

As before, we need an additional equation to determine $a_0$. For now, we use~\eqref{eq:sol11} but treat the Theodorsen function as a constant $C$. The coefficient $a_0$ is then
\begin{equation}
  \label{eq:eig2}
  a_0 = -C(V_0 + V_1) + V_1,
\end{equation}
where $V_k$ is the $k^{\text{th}}$ Chebyshev coefficient of the vertical velocity on the surface of the plate. The $V_k$ are obtained by evaluating the no-penetration condition~\eqref{eq:prob5},
\begin{equation}
  \label{eq:eig3}
  \frac{1}{2}V_0 + \sum_{k=1}^\infty V_k T_k = \frac{1}{2}\dot{\beta_0} + \sum_{k=1}^\infty \left[\dot{\beta_k}T_k + \beta_k T_k'\right].
\end{equation}
Treating $a_0$ in this manner will yield a linear eigenvalue problem. After obtaining the eigenvalues and eigenfunctions of the linear eigenvalue problem, we will use those as initial guesses for the nonlinear eigenvalue problem, which will use the full Theodorsen function. But first, we proceed with the description of the linear eigenvalue problem. 

We can write the equations more compactly as follows:
\begin{align}
  \label{eq:eig4}
  2\tilde{\mathsfbi{R}} \ddot{\boldsymbol{\beta}} + \frac{2}{3}\mathsfbi{D}^2(\tilde{\mathsfbi{S}}{\mathsfbi{D}}^2 \boldsymbol{\beta}) = \boldsymbol{P}, \\
  \boldsymbol{P} = \mathsfbi{A}\boldsymbol{a}, \\
  \mathsfbi{D}\boldsymbol{a} = -\ddot{\boldsymbol{\beta}} - 2\mathsfbi{D}\dot{\boldsymbol{\beta}} - \mathsfbi{D}^2\boldsymbol{\beta}, \\
  \boldsymbol{V} = \dot{\boldsymbol{\beta}} + \mathsfbi{D}\boldsymbol{\beta},
\end{align}
with~\eqref{eq:eig2} for $a_0$. In the above, $\boldsymbol{\beta}$ is a vector of the Chebyshev coefficients of the deflection $Y$, and similarly for $\boldsymbol{P}$ (pressure), $\boldsymbol{a}$ (potential), and $\boldsymbol{V}$ (vertical velocity). $\boldsymbol{P} = \mathsfbi{A}\boldsymbol{a}$ simply states that the Chebyshev coefficients of the pressure are linear combinations of the coefficients $a_k$, and $\mathsfbi{D}$ is the spectral representation of the differentiation operator. Quantities with a tilde over them are spectral representations of multiplication in space, i.e., $\tilde{\mathsfbi{G}} = \mathsfbi{F}\mathsfbi{G}\mathsfbi{F}^{-1}$, where $\mathsfbi{F}$ is the linear operator that maps spatial coordinates to spectral coordinates. 

Putting everything together, we get the following ordinary differential equation:
\begin{eqnarray}
  \label{eq:eig5}
  2\tilde{\mathsfbi{R}}\ddot{\boldsymbol{\beta}} + \frac{2}{3}\mathsfbi{D}^2(\tilde{\mathsfbi{S}}\mathsfbi{D}^2\boldsymbol{\beta}) & = & \mathsfbi{A}[-\mathsfbi{D}^-\ddot{\boldsymbol{\beta}} - 2\mathsfbi{D}^-\mathsfbi{D}\dot{\boldsymbol{\beta}} + \boldsymbol{e_1}(\boldsymbol{e_2} - C\boldsymbol{e_1} - C\boldsymbol{e_2})^T\dot{\boldsymbol{\beta}} \nonumber\\
  &&  - \mathsfbi{D}^-\mathsfbi{D}^2\boldsymbol{\beta} + \boldsymbol{e_1}(\boldsymbol{e_2} - C\boldsymbol{e_1} - C\boldsymbol{e_2})^T \mathsfbi{D}\boldsymbol{\beta}],
\end{eqnarray}
where $\mathsfbi{D}^-$ is the spectral representation of the integration operator that makes the first Chebyshev coefficient zero, and $\boldsymbol{e_k}$ is the $k^{\text{th}}$ Euclidean basis vector. \eqref{eq:eig5} can be written in state-space form as
\begin{equation}
  \label{eq:eig6}
  \left. \begin{array}{ll}
  \displaystyle\frac{\mathrm{d}}{\mathrm{d}t} \begin{bmatrix} \boldsymbol{\beta} \\ \dot{\boldsymbol{\beta}} \end{bmatrix} = \begin{bmatrix} \mathsfbi{0} & \mathsfbi{I} \\ \mathsfbi{M}^{-1}\mathsfbi{A}_1 & \mathsfbi{M}^{-1}\mathsfbi{A}_2 \end{bmatrix} \begin{bmatrix} \boldsymbol{\beta} \\ \dot{\boldsymbol{\beta}} \end{bmatrix}, \\[8pt]
  \mathsfbi{M} = 2\tilde{\mathsfbi{R}} + \mathsfbi{A}\mathsfbi{D}^-, \\[8pt]
  \mathsfbi{A}_1 = -\displaystyle\frac{2}{3}\mathsfbi{D}^2\tilde{\mathsfbi{S}}\mathsfbi{D}^2 - \mathsfbi{A}\mathsfbi{D}^-\mathsfbi{D}^2 + \mathsfbi{A}\boldsymbol{e_1}(\boldsymbol{e_2} - C\boldsymbol{e_1} - C\boldsymbol{e_2})^T \mathsfbi{D}, \\[8pt]
  \mathsfbi{A}_2 = -2\mathsfbi{A}\mathsfbi{D}^-\mathsfbi{D} + \mathsfbi{A}\boldsymbol{e_1}(\boldsymbol{e_2} - C\boldsymbol{e_1} - C\boldsymbol{e_2})^T.
  \end{array}\right\}
\end{equation}

When numerically solving the system, the infinite series are truncated to finite series. In order to incorporate the four boundary conditions into \eqref{eq:eig6}, the last four rows of the differential equation for $\ddot{\boldsymbol{\beta}}$ are replaced by the boundary conditions. The system is then
\begin{equation}
  \label{eq:qui12}
  \frac{\mathrm{d}}{\mathrm{d}t} \begin{bmatrix} \mathsfbi{I} & \mathsfbi{0} \\ \mathsfbi{0} & \mathsfbi{I}_{-4} \end{bmatrix} \begin{bmatrix} \boldsymbol{\beta} \\ \dot{\boldsymbol{\beta}} \end{bmatrix} = \begin{bmatrix} \mathsfbi{0} & \mathsfbi{I} \\ \mathsfbi{M}^{-1}\mathsfbi{A}_1 & \mathsfbi{M}^{-1}\mathsfbi{A}_2 \end{bmatrix} \begin{bmatrix} \boldsymbol{\beta} \\ \dot{\boldsymbol{\beta}} \end{bmatrix},
\end{equation}
where $\mathsfbi{I}_{-4}$ is the identity matrix with the last four diagonal entries being zeros. The last four rows of the right-hand side are replaced by the boundary conditions. We now have a generalized eigenvalue problem to solve for the eigenvalues of the system.

\subsection{Nonlinear eigenvalue problem}
\label{sec:nl}
Having obtained the solution to the linear eigenvalue problem, we use it as an initial guess for the nonlinear eigenvalue problem. The nonlinear eigenvalue problem is obtained by making the ansatz
\begin{equation}
  \label{eq:nl1}
  \left. \begin{array}{ll}
  Y(x,t) = \Real\{\mathrm{e}^{\lambda t}\hat Y(x)\}, \\[8pt]
  \hat Y(x) = \frac{1}{2}\hat\beta_0 + \sum_{k=1}^\infty \hat\beta_k T_k(x).
  \end{array}\right\}
\end{equation}
This is the same as in Appendix~\ref{sec:sol}, except that we allow the exponent $\lambda$ to be any complex number instead of just an imaginary number. Proceeding as in Appendix~\ref{sec:eig}, we arrive at the following equations:
\begin{align}
  \label{eq:nl2}
  2\lambda^2 \tilde{\mathsfbi{R}}\boldsymbol{\hat\beta} + \frac{2}{3}\mathsfbi{D}^2(\tilde{\mathsfbi{S}}\mathsfbi{D}^2 \boldsymbol{\hat\beta}) = \boldsymbol{\hat P}, \\
  \boldsymbol{\hat P} = \mathsfbi{A}\boldsymbol{\hat a}, \\
  \mathsfbi{D}\boldsymbol{\hat a} = -\lambda^2 \boldsymbol{\hat\beta} - 2\lambda \mathsfbi{D} \boldsymbol{\hat\beta} - \mathsfbi{D}^2 \boldsymbol{\hat\beta}, \\
  \boldsymbol{\hat V} = \lambda \boldsymbol{\hat\beta} + \mathsfbi{D} \boldsymbol{\hat\beta}, \\
  \hat a_0 = -C(\lambda)(\hat V_0 + \hat V_1) + \hat V_1,
\end{align}
where the notation is as in Appendix~\ref{sec:eig}. 

Putting everything together, we get the following equation:
\begin{eqnarray}
  \label{eq:nl3}
  2\lambda^2\tilde{\mathsfbi{R}}\boldsymbol{\hat\beta} + \frac{2}{3}\mathsfbi{D}^2(\tilde{\mathsfbi{S}}\mathsfbi{D}^2\boldsymbol{\hat\beta}) & = & \mathsfbi{A}[-\lambda^2 \mathsfbi{D}^-\boldsymbol{\hat\beta} - 2\lambda \mathsfbi{D}^-\mathsfbi{D}\boldsymbol{\hat\beta} + \lambda \boldsymbol{e_1}(\boldsymbol{e_2} - C(\lambda)\boldsymbol{e_1} - C(\lambda)\boldsymbol{e_2})^T\boldsymbol{\hat\beta} \nonumber\\
  &&  - \mathsfbi{D}^-\mathsfbi{D}^2\boldsymbol{\hat\beta} + \boldsymbol{e_1}(\boldsymbol{e_2} - C(\lambda)\boldsymbol{e_1} - C(\lambda)\boldsymbol{e_2})^T \mathsfbi{D}\boldsymbol{\hat\beta}],
\end{eqnarray}
where the notation is as in Appendix~\ref{sec:eig}. Truncating the upper limit of the infinite series to $N$, \eqref{eq:nl3} gives $N+1$ equations for $N+2$ unknowns (the $N+1$ elements of $\boldsymbol{\hat\beta}$ and $\lambda$). We add an equation which normalizes $\boldsymbol{\hat\beta}$ in order to make the system square. As before, the last four equations are replaced by the boundary conditions. We solve for $\boldsymbol{\hat\beta}$ and $\lambda$ using the Newton-Raphson method, using absolute and relative error tolerances $10^{-6}$. For cases where the Newton-Raphson method did not converge, we calculated the solution by looking at a global picture of the determinant of the system and finding its roots. 

We have previously validated our method for calculating eigenvalues of flexible plates with uniform material properties \citep{floryan2018clarifying}. To the best of our knowledge, no prior work has calculated the eigenvalues of plates with non-uniform material properties, giving us nothing to compare to. We note, however, that the same computer code calculates the eigenvalues for plates with uniform and non-uniform material properties, the former merely a special case of the latter.

\subsection{Quiescent fluid}
\label{sec:quiescent}
Consider the case where the plate is immersed in a quiescent fluid, i.e., where the bending velocity is large compared to the fluid velocity. How do the eigenvalues of the system change? To answer this question, we solve the Euler-Bernoulli and Euler equations~\eqref{eq:prob1}--\eqref{eq:prob2} in the limit of large bending velocity. In this limit, the appropriate time scale to use is the bending time scale, which we choose to be $\sqrt{3\langle\rho_s d\rangle L^4/(4 \langle E d^3\rangle)}$. Non-dimensionalizing the solid and fluid equations using the length scale $L/2$ and the bending time scale yields
\begin{equation}
  \label{eq:qui2}
  \left. \begin{array}{ll}
  R^* Y_{tt} + (S^* Y_{xx})_{xx} = \displaystyle\frac{1}{2\langle R\rangle}\Delta p, \\[8pt]
  \nabla \cdot \mathbf{u} = 0, \\[8pt]
  \mathbf{u}_t + \displaystyle\sqrt{\frac{3 \langle R \rangle}{\langle S \rangle}}\mathbf{u}_x = \nabla\phi,
  \end{array}\right\}
\end{equation}
where $R$ and $S$ are as in~\eqref{eq:prob4}, $R^*$ is the spatial distribution (mean 1) of $R$, $S^*$ is the spatial distribution (mean 1) of $S$, and $\phi = p_\infty - p$. In the above, $x$, $t$, $Y$, $\mathbf{u}$, and $p$ are now dimensionless, with the pressure non-dimensionalized by $\rho_f\langle Ed^3\rangle/(3\langle\rho_s d\rangle L^2)$. The limit of a quiescent flow corresponds to $\langle R \rangle / \langle S \rangle \rightarrow 0$, or equivalently $\langle Ed^3 \rangle / \langle \rho_s d \rangle L^2 \gg U^2$, which explicitly puts this limit in terms of velocity scales. For now, we keep all terms and discuss the limit later. Intuitively, large values of the solid-to-fluid mass ratio $\langle R \rangle$ make the fluid dynamics inconsequential to the deflection of the plate (a heavy plate will be unaffected by the surrounding fluid). 

The fluid additionally satisfies the no-penetration condition, stated as
\begin{equation}
  \label{eq:qui4}
  v|_{x \in [-1,1], y=0} = Y_t + \sqrt{\frac{3 \langle R \rangle }{ \langle S \rangle }}Y_x.
\end{equation}
The boundary conditions on the plate are
\begin{equation}
  \label{eq:qui5}
  Y(-1,t) = 0, \quad Y_x(-1,t) = 0, \quad Y_{xx}(1,t) = 0, \quad Y_{xxx}(1,t) = 0.
\end{equation}

We solve for the fluid motion for a given deflection as in Appendix \ref{sec:sol}. Writing the deflection as
\begin{equation}
  \label{eq:qui6}
  Y(x,t) = \frac{1}{2}\beta_0(t) + \sum_{k=1}^\infty \beta_k(t)T_k(x),
\end{equation}
and the components of the complex potential evaluated on the surface of the plate as
\begin{equation}
  \label{eq:qui7}
  \left. \begin{array}{ll}
  \phi(z=x,t) &= \pm\displaystyle\frac{1}{2}a_0(t)\sqrt{\frac{1-x}{1+x}} \pm \sum_{k=1}^\infty a_k(t)\sin k\theta, \\[8pt]
  \psi(z=x,t) &= \displaystyle\frac{1}{2}a_0(t) + \sum_{k=1}^\infty a_k(t)T_k(x),
  \end{array}\right\}
\end{equation}
the pressure difference across the surface of the plate is
\begin{equation}
  \label{eq:qui8}
  \Delta p(x,t) = a_0(t)\sqrt{\frac{1-x}{1+x}} + 2\sum_{k=1}^\infty a_k(t)\sin k\theta.
\end{equation}
The coefficients $a_k$ are obtained by applying the no-penetration condition,
\begin{equation}
  \label{eq:qui9}
  \frac{\partial \psi}{\partial x}|_{y=0} = -\left(\frac{\partial}{\partial t} + \sqrt{\frac{3 \langle R \rangle }{ \langle S \rangle }}\frac{\partial}{\partial x}\right)^2 Y.
\end{equation}
This does not yield $a_0$, which is instead given by the Laplace domain equation
\begin{equation}
  \label{eq:qui10}
  a_0 = -\sqrt{\frac{3 \langle R \rangle }{ \langle S \rangle }}C(V_0 + V_1) + \sqrt{\frac{3 \langle R \rangle }{ \langle S \rangle }}V_1.
\end{equation}
In the limit of a quiescent fluid ($\langle R \rangle / \langle S \rangle \rightarrow 0$), $a_0 \rightarrow 0$. Thus all of the coefficients $a_k$ are determined by~\eqref{eq:qui9}, which itself simplifies since the second term in the parentheses is zero in the limit $\langle R \rangle / \langle S \rangle \rightarrow 0$. We note that in this limit the only fluid force on the plate is the force due to added mass. 

Putting everything together, we get the following ordinary differential equation:
\begin{equation}
  \label{eq:qui11}
  \widetilde{\mathsfbi{R}^*}\ddot{\boldsymbol{\beta}} + \mathsfbi{D}^2(\widetilde{\mathsfbi{S}^*}\mathsfbi{D}^2 \boldsymbol{\beta}) = -\frac{1}{ \langle R \rangle }\mathsfbi{A}\mathsfbi{D}^- \ddot{\boldsymbol{\beta}},
\end{equation}
where $\boldsymbol{\beta}$ is the vector of coefficients $\beta_k$, $\mathsfbi{D}$ is the spectral representation of the differentiation operator, and $\mathsfbi{D}^-$ is the spectral representation of the integration operator that makes the first Chebyshev coefficient zero. Quantities with a tilde over them are spectral representations of multiplication in space, i.e., $\tilde{\mathsfbi{G}} = \mathsfbi{F}\mathsfbi{G}\mathsfbi{F}^{-1}$, where $\mathsfbi{F}$ is the linear operator that maps spatial coordinates to spectral coordinates. The operator $\mathsfbi{A}$ maps the coefficients $a_k$, which are the coefficients of a \emph{sine} series for the pressure, into the corresponding coefficients of a \emph{cosine} series. If $\mathsfbi{T}_s$ is an operator that takes us from the $x$-domain to the sine domain, and $\mathsfbi{T}_c$ is an operator that takes us from the $x$-domain to the cosine domain, then $\mathsfbi{A} = \mathsfbi{T}_c \mathsfbi{T}_s^{-1}$. \eqref{eq:qui11} can be written in state-space form as
\begin{equation}
  \label{eq:qui12}
  \frac{\mathrm{d}}{\mathrm{d}t} \begin{bmatrix} \boldsymbol{\beta} \\ \dot{\boldsymbol{\beta}} \end{bmatrix} = \begin{bmatrix} \mathsfbi{0} & \mathsfbi{I} \\ -\left({\widetilde{\mathsfbi{R}^*}} + \displaystyle\frac{1}{ \langle R \rangle }\mathsfbi{A}\mathsfbi{D}^-\right)^{-1}\mathsfbi{D}^2\widetilde{\mathsfbi{S}^*}\mathsfbi{D}^2 & \mathsfbi{0} \end{bmatrix} \begin{bmatrix} \boldsymbol{\beta} \\ \dot{\boldsymbol{\beta}} \end{bmatrix}.
\end{equation}

When numerically solving the system, the infinite series are truncated to finite series. In order to incorporate the four boundary conditions into~\eqref{eq:qui12}, the last four rows of the differential equation for $\ddot{\boldsymbol{\beta}}$ are replaced by the boundary conditions. This is fine to do since the last four rows read $\ddot{\beta}_k = 0$ due to four applications of the differentiation operator $\mathsfbi{D}$. The system is then
\begin{equation}
  \label{eq:qui12}
  \frac{\mathrm{d}}{\mathrm{d}t} \begin{bmatrix} \mathsfbi{I} & \mathsfbi{0} \\ \mathsfbi{0} & \mathsfbi{I}_{-4} \end{bmatrix} \begin{bmatrix} \boldsymbol{\beta} \\ \dot{\boldsymbol{\beta}} \end{bmatrix} = \begin{bmatrix} \mathsfbi{0} & \mathsfbi{I} \\ -\left({\widetilde{\mathsfbi{R}^*}} + \displaystyle\frac{1}{ \langle R \rangle }\mathsfbi{A}\mathsfbi{D}^-\right)^{-1}\mathsfbi{D}^2\widetilde{\mathsfbi{S}^*}\mathsfbi{D}^2 & \mathsfbi{0} \end{bmatrix} \begin{bmatrix} \boldsymbol{\beta} \\ \dot{\boldsymbol{\beta}} \end{bmatrix},
\end{equation}
where $\mathsfbi{I}_{-4}$ is the identity matrix with the last four diagonals being zeros. The last four rows of the right-hand side are replaced by the boundary conditions. We now have a generalized eigenvalue problem to solve for the eigenvalues of the system.

\section{Some useful formulas}
\label{sec:form}
The following is a collection of useful definitions and formulas from \citet{moore2017fast} for the Chebyshev method employed here. The (interior) Gauss-Chebyshev points are 
\begin{equation}
  \label{eq:form1}
  x_n = \cos\theta_n, \quad \theta_n = \frac{\upi(2n + 1)}{2(N+1)}, \quad \text{for } n=0, 1, \ldots, N.
\end{equation}
Consider a function $f(x)$ interpolated at these points by the polynomial $p_N(x)$ of degree $N$,
\begin{equation}
  \label{eq:form2}
  \left. \begin{array}{ll}
  f(x_n) = p_N(x_n), \quad \text{for } n=0, 1, \ldots, N, \\[8pt]
  P_N(x_n) = \displaystyle\frac{1}{2}b_0 + \sum_{k=1}^N b_k T_k(x).
  \end{array}\right\}
\end{equation}
On the $\theta$-grid this is
\begin{equation}
  \label{eq:form3}
  f(x_n) = \frac{1}{2}b_0 + \sum_{k=1}^N b_k \cos k\theta_n, \quad \text{for } n=0, 1, \ldots, N.
\end{equation}
Thus we may use the fast discrete cosine transform to transform between a function's values on the collocation points, $f(x_n)$, and the Chebyshev coefficients $b_k$. 

The antiderivative of $p_N(x)$ is
\begin{equation}
  \label{eq:form4}
  \left. \begin{array}{ll}
  D^{-1}p_N(x) = \displaystyle\frac{1}{2}B_0 + \sum_{k=1}^{N+1} B_k T_k(x), \\[8pt]
  B_k = \displaystyle\frac{1}{2k}(b_{k-1} - b_{k+1}), \quad \text{for } n=1, 2, \ldots, N.
  \end{array}\right\}
\end{equation}
Here, $B_0$ is a free constant of integration. 

The derivative of $p_N(x)$ is
\begin{equation}
  \label{eq:form5}
  \left. \begin{array}{ll}
  Dp_N(x) = \displaystyle\frac{1}{2}b_0' + \sum_{k=1}^{N} b_k' T_k(x), \\[8pt]
  b'_{N+1} = b'_N = 0, \\[8pt]
  b'_k = b'_{k+2} + 2(k+1)b_{k+1}, \quad \text{for } n=N-1, N-2, \ldots, 0.
  \end{array}\right\}
\end{equation}

Since the endpoints $x = \pm 1$ are not part of the collocation grid, we give a formula to evaluate the function at the endpoints,
\begin{equation}
  \label{eq:form6}
  p_N(\pm 1) = \frac{1}{2}b_0 + \sum_{k=1}^N (\pm 1)^k b_k.
\end{equation}

\bibliographystyle{jfm}
\bibliography{references}

\end{document}